\pdfoutput=1
\newif\ifSC

\ifSC
\documentclass[onecolumn,draftclsnofoot,12pt]{IEEEtran}
\else
\documentclass[10pt,twocolumn]{IEEEtran}
\fi
\usepackage{amssymb}
\usepackage{pifont}

\usepackage{bbm}
\usepackage{amssymb}
\usepackage{amsmath,amsthm}
\usepackage{color}
\usepackage{bm}
\usepackage{amsfonts}
\usepackage{graphicx}
\usepackage{verbatim}
\usepackage{epstopdf}
\usepackage{cite}
\usepackage{tabulary}
\usepackage{mathtools}
\usepackage[caption=false]{subfig}
\usepackage{siunitx}
\usepackage{enumitem}
\usepackage{booktabs}
\usepackage{array}
\usepackage{algorithm}

\usepackage{algorithmic}

\newcommand{\bmx}{\bm{x}}

\newcommand{\rmh}{{\rm h}}
\newcommand{\rmv}{{ \rm v}}
\newcommand{\rms}{{ \rm s}}
\newcommand{\tcal}{\mathcal{T}}
\newcommand{\trp}{P_{\rm t}}
\newcommand{\1}{\mathbbm{1}}

\newcommand{\PGF}{\mathcal{P}}

\newcommand{\elv}{\mathrm{e}}

\newcommand{\dv}{\mathrm{d}}

\newcommand{\expS}[1]{\exp{\left(#1\right)}}

\def\home{\hbox{\kern3pt \vbox to13pt{}%
   \pdfliteral{q 0 0 m 0 5 l 5 10 l 10 5 l 10 0 l 7 0 l 7 5 l 3 5 l 3 0 l f
               1 j 1 J -2 5 m 5 12 l 12 5 l S Q }%
   \kern 13pt}}

\newcommand{\x}{\mathbf{x}}
\newcommand{\y}{\mathbf{y}}
\newcommand{\bt}{\mathbf{b}}
\newcommand{\ob}{\mathrm{o}}
\newcommand{\sfx}{\mathsf{x}}
\newcommand{\sfy}{\mathsf{y}}
\newcommand{\trm}{\mathrm{t}}
\newcommand{\SINR}{\mathrm{SINR}}

\newcommand{\prm}{\mathrm{p}}
\newcommand{\crm}{\mathrm{c}}

\newtheorem{theorem}{Theorem}

\newtheorem{definition}{Definition}
\newtheorem{corollary}{Corollary}[theorem]
\newtheorem{remark}{Remark}
\usepackage{tikz}
\usepackage{booktabs}
\usepackage{tabularx}
\usepackage[caption=false]{subfig}
\usepackage{color}            

\usetikzlibrary{calc,angles,quotes}

\ifCLASSINFOpdf
 
\else
\fi

\allowdisplaybreaks

\begin{document}

\title{Modeling and Analysis of Sensing Assisted UAV Networks for Urban Vehicular Communications}
\author{Kaushlendra~Pandey, Nithin V Sabu, Abhishek K. Gupta
	\thanks{Kaushlendra~Pandey is with the Department
		of Electronics and Communication Engineering, Central Institute of Technology Kokrajhar, India e-mail: (kk.pandey@cit.ac.in).}
	\thanks{Nithin V Sabu is with the National Institute of Technology Calicut, Kozhikode, Kerala 673601, India (Email:{nithinvs@nitc.ac.in}).} \thanks{ A. K. Gupta is with Indian Institute of Technology Kanpur, Kanpur UP 208016, India (Email:gkrabhi@iitk.ac.in).}}


%



\maketitle

\begin{abstract}
Urban vehicular networks (VNs) demand seamless connectivity and situational awareness within road-constrained environments, motivating the deployment of unmanned aerial vehicles (UAVs) platforms capable of simultaneously sensing vehicles and establishing communication with them. In this paper, we present a sensing-assisted UAV network that provides connectivity to the vehicles in an urban area. The road network of the urban area is modeled as Manhattan Poisson line process (MPLP), and the random location of vehicles on each road is modeled as one dimensional Poisson point processes (PPPs).
UAVs are distributed in the urban area at a fixed altitude and provide connectivity after sensing the vehicles. Their locations are modeled as a two-dimensional homogeneous PPP. Combined with the fixed altitude, this results in a three-dimensional spatial configuration. We incorporate an elevation dependent blockage model and define the sensing radius based on detection probability (DP), showing that it is jointly limited by signal strength and blockage effects.
We derive the DP and characterize the typical UAV's sensing region within the reliability requirements. We also derive the Laplace transform (LT) of aggregate interference accounting for directional patterns and sensing-driven activity, and analyze the resulting coverage probability (CP). Finally, we obtain the rate coverage (RC) of sensed vehicles falling within the UAV's sensing zone. Numerical results shows that increasing altitude degrades sensing and coverage performance, whereas RC exhibits a non-monotonic trend, first decreasing and then increasing with altitude.
\end{abstract}

\IEEEpeerreviewmaketitle

\section{Introduction}
{{Emerging urban 6G vehicular networks (VNs) supporting applications such as 3D map navigation, real-time traffic situational awareness, and cooperative driving, essentially demands reliable connectivity.
The current infrastructure network is inadequate to reliably support these services, as high vehicular mobility and severe urban blockage conditions significantly degrade link reliability. To address these connectivity challenges low-altitude UAVs flying below 500 \si{feet} have gained significant attention as a key architectural component of future 6G-enabled VNs \cite{shi2018drone,UAVassisted,huang2024_low,jiang2024_6g_ntn_lae}. In particular, UAV-assisted VNs support high vehicular mobility by establishing reliable line-of-sight (LoS) links due to their flexible deployment, aerial mobility, and directional connectivity \cite{alhourani2014modeling,wu2018,deng2019joint,He2019AntiInterference,Oubbati2019ECaD}. Moreover, beyond serving as aerial base stations or relays, next-generation UAVs are expected to support integrated sensing and communication (ISAC), enabling simultaneous detection, localization, and communication with vehicles \cite{Hu2026MADQN,Pang2024DynamicISAC}. Recent advances in ISAC have highlighted the potential benefits of waveform co-design, beamforming, and spectrum sharing between sensing and communication tasks (see, e.g., \cite{Pang2024DynamicISAC,Liu2022ISAC,Shen2021ISAC}). In particular, sensing-assisted communication frameworks, in which sensing information is leveraged to adapt link design, are especially suitable for blockage-dominant THz 6G vehicular networks, where high mobility and severe propagation losses pose significant challenges \cite{armeniakos2025stochastic}. These coupled effects are absent in conventional aerial base station models and remain insufficiently characterized in existing studies, thereby motivating a systematic analytical treatment tailored to urban vehicular low altitude wireless network (LAWN) deployments.}}
\subsection{Related work}
 UAV-assisted vehicular networks (VNs) build on the broader foundation of UAV-enabled wireless communications. Comprehensive surveys have examined UAVs functioning as aerial base stations, relay nodes, and cellular users, characterizing their coverage, reliability, and latency performance \cite{zeng2019accessing,Zhang2019MmWaveUAV,Mozaffari2019,9508366}. These studies consistently highlight that downlink connectivity from aerial platforms is challenged by altitude-dependent channel characteristics, interference, and resource constraints. Unlike terrestrial base stations, UAVs offer flexible on-demand deployment and altitude-adaptive positioning, which are particularly valuable for serving vehicular users whose high mobility and unpredictable spatial distribution strain fixed ground infrastructure \cite{telikani2025unmanned}. Within this context, recent work has addressed specific aspects of UAV deployment for vehicular connectivity. In \cite{he2023aerial}, the coverage radius of UAV base stations was maximized by optimizing the UAV altitude, while \cite{zhang2024performance} analyzed a UAV-assisted VN supporting both UAV-to-vehicle and vehicle-to-vehicle links, determining near-optimal UAV altitude to maximize the number of served users. An air-ground integrated VN was studied in \cite{jiang2024performance}, with emphasis on how UAV positioning affects task offloading performance. However, these works treat UAVs purely as communication nodes and do not exploit sensing information to improve link establishment or beam alignment.
    \begin{table*}[!t]
  	\centering
  	\caption{Comparison of the proposed framework with related works.}
  	\label{tab:comparison}
  	\renewcommand{\arraystretch}{1.15}
  	\footnotesize
  	\begin{tabular}{|l|c|c|c|c|c|c|c|c|}
  		\hline
  		\textbf{Reference} & \textbf{UAV} & \textbf{ISAC/} & \textbf{Road} & \textbf{Stochastic} & \textbf{Blockage} & \textbf{Detection} & \textbf{Coverage/Rate} & \textbf{Beamwidth} \\
  		& \textbf{Platform} & \textbf{Sensing} & \textbf{Topology} & \textbf{Geometry} & \textbf{Model} & \textbf{Prob.} & \textbf{Analysis} & \textbf{Trade-off} \\
  		\hline
  		\cite{he2023aerial} & \checkmark & \ding{55} & \ding{55} & \ding{55} & \ding{55} & \ding{55} & \ding{55} & \ding{55} \\
  		\hline
  		\cite{zhang2024performance} & \checkmark & \ding{55} & \ding{55} & \ding{55} & \ding{55} & \ding{55} & \checkmark & \ding{55} \\
  		\hline
  		\cite{UAVassisted} & \checkmark & \ding{55} & 1D PPP & \checkmark & \ding{55} & \ding{55} & \checkmark & \ding{55} \\
  		\hline
  		\cite{chetlur2018coverage} & \ding{55} & \ding{55} & PLP-PPP & \checkmark & \ding{55} & \ding{55} & \checkmark & \ding{55} \\
  		\hline
  		\cite{31} & \checkmark & \checkmark & \ding{55} & \ding{55} & \ding{55} & \ding{55} & \checkmark & \ding{55} \\
  		\hline
  		\cite{32} & \checkmark & \checkmark & \ding{55} & \ding{55} & \ding{55} & \ding{55} & \checkmark & \checkmark \\
  		\hline
  		\cite{33} & \checkmark & \checkmark & \ding{55} & \ding{55} & \ding{55} & \ding{55} & \checkmark & \ding{55} \\
  		\hline
  		\cite{Pang2024DynamicISAC} & \checkmark & \checkmark & Straight & \ding{55} & \ding{55} & \ding{55} & \checkmark & \checkmark \\
  		\hline
  		\textbf{This work} & \checkmark & \checkmark & MPLP & \checkmark & \checkmark & \checkmark & \checkmark & \checkmark \\
  		\hline
  	\end{tabular}
  \end{table*}
  In vehicular environments, especially at millimeter-wave and higher frequencies, establishing and maintaining directional links with fast-moving vehicles is extremely challenging. Sensing-assisted communication addresses this by using radar-derived information such as vehicle location, speed, and direction to align beams proactively, eliminating the overhead of conventional beam training. This paradigm is central to ISAC, where the same hardware and spectrum resources are shared between radar sensing and data transmission.  Several studies have explored sensing integration into UAV platforms. In \cite{31}, a UAV-enabled ISAC framework jointly optimized trajectory and target state estimation to balance sensing accuracy with communication throughput. The work in \cite{32} optimized transmit beamforming for both quasi-stationary and mobile UAV scenarios under sensing constraints, while \cite{33} introduced a periodic sensing-and-communication protocol for adaptive power and frequency allocation. For vehicular settings, \cite{Pang2024DynamicISAC} developed dynamic beamforming and trajectory optimization for UAV-vehicle tracking. While these contributions advance UAV-ISAC design, they rely on optimization-based approaches for specific configurations and lack the scalable analytical characterization needed for system-level performance evaluation across diverse urban deployments. To enable such scalable analysis, stochastic geometry (SG) has emerged as the primary mathematical framework for wireless network modeling. Line processes such as the Poisson line process and the Manhattan Poisson line process (MPLP) provide tractable models for random and grid-structured urban road networks \cite{chetlur2020shortest,dhillon2022poisson}. Building on this foundation, \cite{chetlur2018coverage} modeled vehicular nodes as one-dimensional Poisson point processes distributed along roads generated by a Poisson line process, forming a Cox process, and derived coverage probability (CP) expressions. In \cite{UAVassisted}, UAV-assisted VNs were investigated using a simplified one-dimensional PPP model that does not capture urban road structure.\\
 \noindent {Despite these contributions, several critical gaps remain. No existing framework jointly models UAV sensing and communication in urban vehicular environments with structured road networks. The role of radar-derived location data in enabling proactive beam alignment eliminating separate beam training, has not been analytically characterized. The altitude-dependent interplay among sensing reliability, coverage, and rate performance remains understudied, despite altitude simultaneously affecting path loss, LoS probability, and the impact of antenna beamwidth on coupled sensing-communication performance requires rigorous analysis. These gaps motivate the SG-based framework developed here, which captures urban road topology, models altitude-dependent sensing with blockage, and derives metrics revealing the fundamental trade-offs in sensing-assisted UAV networks for urban vehicular communications.}
\subsection{Contributions:} This paper develops an analytical framework for sensing-assisted LAWN serving urban VNs. We quantify fundamental 
trade-offs between sensing reliability, CP, and interference. The main contributions are as follows.

\begin{itemize}
	\item An analytical framework is employed where roads are modeled as a MPLP, vehicles as $1$D Poisson point processes along roads, and UAVs as a $2$D Poisson point process. This captures the structured geometry of urban VNs and represents the analytical framework combining road-constrained vehicular distributions with aerial ISAC nodes.
	
	\item The detection probability under directional radar sensing with altitude-dependent path-loss is derived. The sensing radius is shown to be limited by either SNR or LoS blockage, identifying two distinct operating regimes. The distribution of the nearest detected vehicle distance and the UAV activation probability are characterized, with an asymptotic analysis showing that in high-traffic environments the activation probability saturates, revealing that UAV-assisted connectivity is ultimately limited by sensing geometry rather than vehicular density.
	
	\item The LT of aggregate interference is derived while accounting for stochastic UAV deployment, directional antenna patterns, and sensing-driven transmitter activity. This enables interference characterization under realistic sensing-assisted communication  operation where UAV activity is coupled to sensing success. The CP is analyzed by conditioning on successful detection of the nearest vehicle, and the RC is derived accounting for multi-user load distribution. The framework quantifies the coupled impact of altitude, beamwidth, vehicular density, and UAV deployment density on communication performance.
	
	\item Numerical results validate the analytical expressions and reveal fundamental trade-offs.  The results provide system-level insights for the design of antenna patterns, deployment density, and altitude selection in sensing-assisted LAWN for urban VNs.
\end{itemize}
\textit{Notation:}
	Notations used in this paper are summarized in
	notation Table~\ref{table:notation}. Here, we introduce only those notational definitions and transformations that
	cannot be concisely presented in the notation table. Each line $\ell$ parallel to the $\sfx$- or $\sfy$-axis is uniquely identified
	by its perpendicular distance $\rho$ from the origin. The base of a line
	$\ell$ is defined as the point on $\ell$ closest to the origin. For a fixed $\rho \in \mathbb{R}$ and a one-dimensional coordinate
	$\bmx \in \mathbb{R}$, the transformations
	\begin{align}
		\tcal_\sfx(\bmx) &= (\bmx,\rho), \quad
		\tcal_\sfy(\bmx) = (\rho,\bmx),
	\end{align}
	map a one-dimensional location onto horizontal and vertical lines in
	$\mathbb{R}^{2}$, respectively. These mappings are used to embed
	one-dimensional point processes into the two-dimensional plane. 	For a non-negative integer-valued random variable $X$ with probability generating function (PGF) $\mathcal{P}_{X}(s)$, we denote its first and second derivatives with respect to $s$ as $\mathcal{P}^{(1)}_{X}(s)$ and $\mathcal{P}^{(2)}_{X}(s)$, respectively.	The summation notation $\sum_{\mathrm{N}_{k}}$ denotes summation over all
	$k$-tuples $(b_1,b_2,\ldots,b_m)$ of non-negative integers satisfying
	$\sum_{i=1}^{m} i\,b_i = m$. This notation arises in higher-order derivative
	expansions of generating functions. For example, $k=0$, ${\rm N}_{0}=\{(0)\}$, $k=1$, ${\rm N}_1=\{(1)\}$, and $k=2$, ${\rm N}_2=\{(2,0),(0,1)\}.$

\begin{table}[!t]
	\renewcommand{\arraystretch}{1.15}
	\centering
	\small
		\caption{Notation table}
	\label{table:notation}
	\begin{tabularx}{\columnwidth}{lX}
		\toprule
		\textbf{Symbol} & \textbf{Description} \\
		\midrule
		$\Phi_\ell$ & Road network modeled as MPLP with density $\lambda_\ell$ \\
		$\Psi$ & Vehicle locations modeled as MPLP–PPP \\
		$\Phi_u$ & UAV locations  as PPP with density $\lambda_u$ \\
		$H$ & UAV altitude \\
		$\gamma$ & Antenna beamwidth \\
		$\mathbf{x}=(r,\theta)$ & Vehicle location $\mathbf{x}\in\mathbb{R}^2$ in polar coordinates \\
		$d(r)$ & 3D UAV–vehicle distance $\sqrt{r^2 + H^2}$ \\
		$r_{\mathrm{s}}, d_{\mathrm{s}}$ & Horizontal and 3D sensing radii \\
		{$r_{\rm L}$} & {Blockage-limited horizontal radius} \\
		{$r_{\rm s}^{\rm snr}$} & {SNR-limited horizontal sensing radius} \\
		$\alpha, \kappa_f$ & Path-loss exponent and absorption coefficient \\
		$G_\rms(\gamma)=G_\crm(\gamma)$ & {Sensing/communication main-lobe gain} \\
		$\sigma, \bar{\sigma}$ & Radar cross section and mean value \\
		$\tau_{\mathrm{s}}, \delta$ & Sensing SNR and reliability thresholds \\
		${\rm P_d}(r)$ & DP at distance $r$ \\
		$N_0, \tau_c$ & Noise power and SINR threshold \\
		${\rm p_a}$ & UAV activation probability \\
		$N(r_{\mathrm{s}})$ & Number of vehicles in sensing zone \\
		$R_k$ & Distance to the $k$-th nearest vehicle \\
		${\rm P}_c^k(\tau_c)$ & CP of the $k$-th vehicle \\
		$r_c(\tau)$ & Rate CP \\
		$p_{\sf L}(r,H)$ & LoS probability at distance $r$ and altitude $H$ \\
		$\Lambda_{\rm h}(\rho), \Lambda_{\rm v}(\rho)$ & LoS-thinned vehicle counts on chord \\
		\bottomrule
	\end{tabularx}

\end{table}

{\begin{definition}[MPLP]
Let there be two independent $1$D PPPs, each with density
$\lambda_{\ell}$, defined on the representation space
$\mathbb{C}\equiv\mathbb{R}\times\{0,\pi/2\}$.
The points with orientation $0$ generate horizontal lines forming
$\Phi_{\rm h}$, while the points with orientation $\pi/2$ generate vertical
lines forming $\Phi_{\rm v}$ in $\mathbb{R}^{2}$.
Taking the union of these two-line processes,
\begin{align}
	\Phi_{\ell}=\Phi_{\rm h}\cup \Phi_{\rm v},
\end{align}
constitutes a MPLP.

\end{definition}\noindent In MPLP, the number of horizontal (respectively, vertical) lines
intersecting a convex region $K$ is Poisson distributed with mean
$\lambda_{\ell}\,{\sf P}(K)$, where ${\sf P}(K)$ denotes the projection length of region
$K$ onto the $\sfy$-axis (respectively, the $\sfx$-axis).
\begin{definition}[MPLP-PPP]
	Let $\{\psi^{\rm h}_i,\; i\in\mathbb{N}\}$ and $\{\psi^{\rm v}_j,\; j\in\mathbb{N}\}$ be two independent collections of i.i.d.\ $1$D PPPs on $\mathbb{R}$, each with density
	$\lambda$. The $i$th PPP $\psi^{\rm h}_i$ is independently assigned to the $i$th horizontal line $\ell_{i}\in\Phi_{\rm h}$, and the $j$th PPP $\psi^{\rm v}_j$ is independently assigned to the $j$th vertical line
	$\ell_{j}\in\Phi_{\rm v}$. Precisely, the mapped point processes are defined as
	\begin{align*}
		&\Psi_{\rm h,\ell_{i}}
		=
		\bigcup\nolimits_{\bm{x}\in\psi^{\rm h}_{i}}
		\bigl\{ \mathbf{x}=\mathcal{T}_{\sfx}(\bm{x}) \bigr\},
		\,\,
		\Psi_{\rmv,\ell_{j}}
		=
		\bigcup\nolimits_{\bm{x}\in\psi^{\rm v}_{j}}
		\bigl\{ \mathbf{x}=\mathcal{T}_{\sfy}(\bm{x}) \bigr\}.
	\end{align*}
	The superposition of the transformed point processes on all horizontal and
	vertical lines yields
	\begin{align*}
		\Psi_{\rm h}=\bigcup\nolimits_{\ell_{i}\in\Phi_{\rm h}}\Psi_{\rm h,\ell_{i}},\quad
		\Psi_{\rm v}=\bigcup\nolimits_{\ell_{j}\in\Phi_{\rm v}}\Psi_{\rmv,\ell_{j}},
	\end{align*}
	and the resulting point process $	\Psi=\Psi_{\rm h}\cup\Psi_{\rm v}$ is called the MPLP-PPP.	
\end{definition}}
\noindent The remainder of this paper is organized as follows. Section II presents the system model, including road network modeling, UAV deployment, directional sensing, and communication models. Section III analyzes the 
sensing performance, deriving the DP, sensing radius, and UAV activation probability. Section IV presents the communication performance analysis, including interference characterization and coverage 
probability. Section V provides numerical results and insights. Section VI concludes the paper with future directions. 
\section{System Model}
{We consider an urban scenario where a fleet of UAVs is deployed at a common altitude~$H$ to provide downlink connectivity to ground vehicles. {Each UAV operates sector-by-sector, it steers its directional beam to a sector, performs radar sensing to detect vehicles, and immediately communicates with any detected vehicles in that sector before advancing to the next. This {sense-then-communicate} cycle repeats across all $2\pi/\gamma$ sectors spanning the full $360^{\circ}$ azimuth.} UAVs that detect no vehicle {in any sector during the entire {cycle}} remain idle and do not transmit, thereby reducing network interference. The key analytical objective is to characterize how the interplay among UAV altitude, antenna beamwidth, urban blockage, and road geometry governs the sensing reliability, UAV activation probability, and the resulting communication coverage and rate. In what follows, we present the spatial models for roads, vehicles, and UAVs, followed by the sensing model, the communication model, and the performance metrics.
\subsection{Road network and vehicle model} The urban road network is 
modeled as an MPLP $\Phi_\ell$ with line density $\lambda_\ell$.
{MPLP accurately captures the Manhattan road structure, where roads run in two perpendicular directions forming the rectangular layout typical of urban cities. The MPLP model has been validated against real
	urban road networks \cite{dhillon2022poisson}. This captures the road-constrained mobility of vehicles while preserving spatial randomness.}} Vehicles on the roads of $\Phi_\ell$ are 
modeled as an MPLP-PPP $\Psi$, where vehicles on each 
road follow an independent $1$D PPP with density $\lambda$. Each vehicle acts as a potential radar target for the UAVs. During the sensing phase, the UAV illuminates the ground with a directional radar beam to detect the {presence} and estimate the {angular location} of vehicles on the roads below. Upon detection, the UAV {immediately} reuses the beam alignment to establish a communication link, naturally coupling sensing and communication.
\subsection{UAV deployment} {The ground-projected locations of UAVs are modeled as a homogeneous $2$D PPP $\Phi_u = \{\y_i,\, i \in \mathbb{N}\}$ with density $\lambda_u$. The PPP model captures the spatial randomness inherent in on-demand UAV dispatches, where no predetermined positions are available. Assuming a fixed altitude $H$, consistent with low-altitude airspace regulations, allows $H$ to serve as a single tunable design parameter, consequently, the $3$D UAV--vehicle geometry is fully determined by the ground projection and altitude. {Each UAV~$\y_i$ is assigned an independent mark~$\Theta_i\sim\mathrm{Unif}[0,2\pi)$ representing its {starting scan angle}, so that
$\tilde{\Phi}_u=\{(\y_i,\Theta_i)\}$ is an independently marked PPP \cite{haenggibook} Once~$\Theta_i$ is drawn, the UAV's sector grid is fixed and it scans sectors $\Theta_i,\,\Theta_i+\gamma,\,\Theta_i+2\gamma,\,\ldots$ deterministically. Thus $\Theta_i$ is the random rotational offset of the sector grid, and different UAVs have independently oriented grids. Without loss of generality the typical UAV at the origin we write $\Theta_{\ob}$, for an interfering UAV at~$\y$ we write~$\Theta_{\y}$.} }
\subsection{Sensing model} Each UAV is equipped with a monostatic radar operating at carrier frequency~$f_c$ that uses a directional beam of angular width~$\gamma$ to detect ground vehicles. Because the beam illuminates only a fraction of the azimuth at any instant, the UAV performs a sequential angular scan in $2\pi/\gamma$ discrete steps covering the full $360^{\circ}$. {Starting from~$\Theta_{\ob}$, the typical UAV visits $\Theta_{\ob,j} = \Theta_{\ob}+j\gamma$ for $j = 0, 1, \ldots, 2\pi/\gamma - 1$.} In each step the boresight points in a fixed direction~${\Theta_{\ob,j}}$, and vehicles within the angular sector $[{\Theta_{\ob,j}}-\gamma/2,\,{\Theta_{\ob,j}}+\gamma/2]$ and within the maximum sensing range~$r_{\rm s}$ are potential detection targets. {We refer to the region covered by a single step, the intersection of the disk of radius~$r_{\rm s}$ and the angular sector, as the {instantaneous sensing zone}. Since the full scan covers $2\pi$, the sensing analysis (DP, sensing radius, activation probability) does not depend on the particular value of~$\Theta_{\ob}$.} We now describe the target fluctuation model, the path-loss model, the blockage model, and the antenna gain model that together determine the detection performance.
\subsection{Swerling I target model} Each ground vehicle acts as a radar target whose echo signal fluctuates due to the vehicle's complex shape and varying aspect angle relative to the UAV.
We model these fluctuations using the Swerling~I model \cite{swerling2003probability}, in which the radar cross section (RCS) $\sigma$ follows an exponential distribution with mean~$\bar{\sigma}$, i.e., $f_{\sigma}(\sigma)=\frac{1}{\bar{\sigma}}e^{-\sigma/\bar{\sigma}}$ for $\sigma\geq 0$. The exponential distribution arises from modeling the vehicle as a collection of many small, independently phased scatterers--a standard assumption in radar theory \cite{Persson2017RadarRCS,Fang2020StochasticRadarRCS}. Under Swerling~I, the RCS remains constant during single scan but varies independently across scans. {Importantly, the RCS distribution captures the fluctuation statistics of the target reflectivity and is {independent} of the target's distance from the UAV, the distance-dependent signal decay is captured separately by the path-loss model described next.}
{A unified transmit power~$\trp$ is used for both sensing and communication. The large-scale signal attenuation between a UAV and a ground vehicle is modeled as a combination of geometric spreading and frequency-dependent molecular absorption. At the millimeter-wave carrier frequency considered here ($f_c = 60$~GHz), oxygen molecules along the propagation path introduce absorption losses of approximately $15$~\si{dB/km} \cite{ITU-R-P676}, at lower carrier frequencies this term is negligible. Accordingly, the path-loss model is expressed as
\begin{align}
	g(r)= r^{-\alpha}\exp\!\left(-\kappa_f r\right),
\end{align}where $\alpha$ denotes the path-loss exponent and $\kappa_f$ represents the molecular absorption coefficient at the operating carrier frequency. \\
Fig.~\ref{fig:systemmodelfinal} illustrates the network layout. UAVs hover above a grid of urban roads on which vehicles are distributed. An {active} UAV is one that has detected at least one vehicle within its {sensing zone of radius~$r_{\rm s}$ during a full angular scan and is engaged in directional downlink communication. An {inactive} UAV has completed a full scan without detecting any vehicle and therefore does not transmit. 
{\subsection{Blockage model} In UAV-to-vehicle communication, the probability of LoS depends on the elevation angle $\theta_\elv=\arctan(H/r)$, where $H$ is the altitude of the UAV and $r$ is the horizontal distance of the vehicle from the UAV. Unlike terrestrial blockage, which is primarily distance-dependent, UAV links experience a different blockage regime: the region directly below the UAV is almost certainly LoS, while at shallow elevation angles buildings increasingly obstruct the path. We adopt the widely used sigmoid LoS model \cite{al2014optimal}
		\begin{equation}\label{eq:PLoS}
			{ p}_{\sf L}(r,H) = \frac{1}{1 + a \exp\!\left(-b\!\left[\frac{180}{\pi}\arctan\!\left(\frac{H}{r}\right) - a\right]\right)},
		\end{equation}
		where $a$ and $b$ are environment-dependent parameters. The NLoS probability is simply
		\begin{equation}
			p_{\sf N}(r,H) = 1 - p_{\rm L}(r, H).
\end{equation}

\noindent\textit{Step-like nature of the blockage probability:} A key property of the sigmoid LoS model in \eqref{eq:PLoS} is its step-like transition behavior as a function of the horizontal distance~$r$ for a fixed altitude~$H$. For small~$r$ (i.e., large elevation angles), the LoS probability $p_{\sf L}(r,H) \approx 1$, implying near-certain LoS. As $r$ increases beyond a critical distance, $p_{\sf L}(r,H)$ drops sharply toward zero, exhibiting an approximate step-function transition. The steepness of this transition is governed by the parameter~$b$: larger values of~$b$ produce a sharper drop, making the LoS probability resemble a unit step function.  Formally, in the limit $b \to \infty$, the LoS probability reduces to a step function:
\begin{equation}\label{eq:PLoS_step}
	\lim_{b\to\infty} p_{\sf L}(r,H) = \begin{cases}
		1, & \text{if } \arctan(H/r) > \frac{\pi a}{180},\\
		0, & \text{if } \arctan(H/r) < \frac{\pi a}{180},
	\end{cases}
\end{equation}
which defines a blockage-limited radius $r_{\rm L} = H/\tan(\pi a/180)$ beyond which LoS is effectively lost. This step-like behavior implies that beyond a certain horizontal range, the LoS path is almost entirely blocked by urban structures, regardless of the transmit power or antenna gain. Consequently, the effective sensing and communication range of the UAV is fundamentally limited by the urban blockage environment, not merely by signal attenuation.

\noindent\textit{Use of the blockage model:} We employ the LoS model at two levels of approximation. For defining the blockage-limited sensing radius, we adopt the step-function limit in~\eqref{eq:PLoS_step}, yielding the closed-form cutoff $r_{\rm L}=H/\tan(\pi a/180)$. This is appropriate because the sensing radius is a hard design boundary, a vehicle is either within the reliable sensing range or not. For the communication analysis (interference characterization and CP), we retain the full sigmoid model~\eqref{eq:PLoS} because interference is a statistical aggregate over all UAVs, and the smooth LoS transition accurately weights each interferer's contribution.
  }
  \subsection{Directional antenna gain model}
  	The sensing antenna is modeled using a sectored beam approximation with beamwidth
  	$\gamma$, which is widely adopted in ISAC enable LAWN analysis to capture main-lobe characteristics while maintaining analytical tractability
  	\cite{32,31,33}.
  	As shown in Fig.~\ref{fig:systemmodelantenna}, for the typical UAV at the
  	origin and a vehicle at $\mathbf{x}=(r,\theta)$, the effective sensing gain
  	is
  	\begin{align}\label{eq:Geff}
  		G_{\mathrm{eff}}({\Theta_{\ob}})
  		=
  		\begin{cases}
  			G_{\rms}(\gamma)={2\pi}/{(3\gamma)},
  			& \text{if }\ |\theta-{\Theta_{\ob}}|\le \gamma/2, \\[1.2ex]
  			0,  & \text{otherwise},
  		\end{cases}
  	\end{align}
  	where ${\Theta_{\ob}}$ denotes the current boresight direction {of the typical UAV}.
The	main-lobe gain satisfies $G\approx k/\gamma$, where the constant $k=120^{\circ}= 2\pi/3$ corresponds to the standard approximation for a linear aperture \cite{balanis2016, skolnik2008}.
  This sectored gain model is consistent with recent stochastic-geometry
  analysis of ISAC networks \cite{nabil2024beamwidth, ram2025radar}.
Since $G_{\rms}(\gamma)$ is a deterministic function of~$\gamma$,
it scales the SNR by a fixed constant without affecting the
independence structure, preserving analytical tractability.
  } For simplicity, we assume that the sidelobe gain is negligible, consistent with directional beamforming systems 
where the main-to-side lobe ratio exceeds 20~\si{dB} \cite{liu2018mu}. We note that the antenna model captures only the azimuthal beamwidth~$\gamma$, the elevation beampattern is assumed to be sufficiently broad to illuminate the ground at all relevant horizontal distances. 
\begin{figure}[t!]
	\centering
	\includegraphics[width=\linewidth]{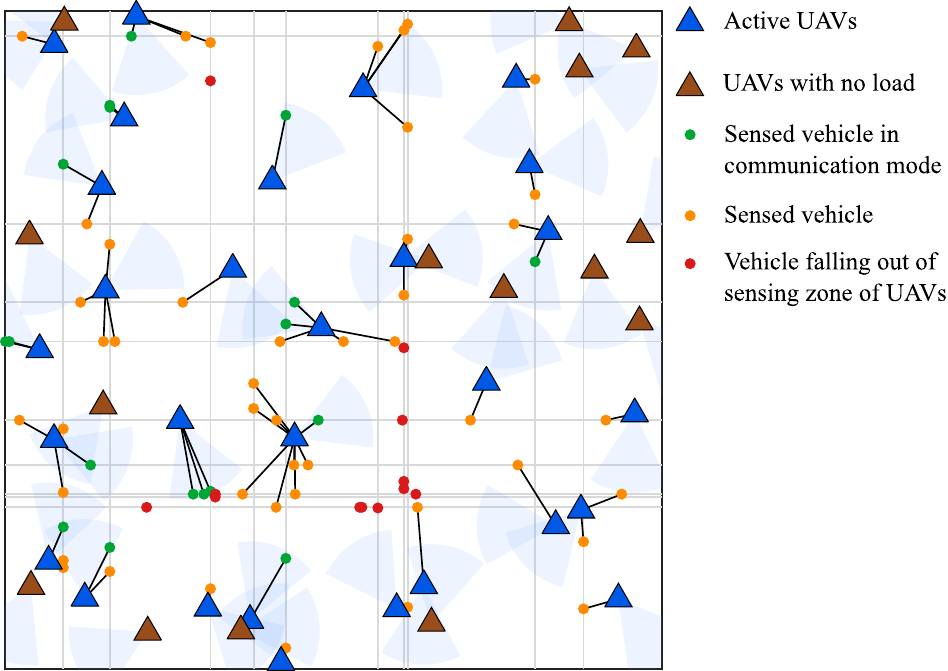}
	\caption{System model illustrating active UAVs with their sensing zone, inactive UAVs, vehicles on the road.}
	\label{fig:systemmodelfinal}
\end{figure}
  
\noindent	Without loss of generality, we place the typical UAV at the origin, i.e., $\ob \in \Phi_u$, and analyze the sensing and communication performance from its perspective. For a vehicle located at ${\bf x}=(r,\theta)\in\mathbb{R}^2$ relative to the UAV's ground projection, the radar
	echo-based sensing signal-to-noise ratio (SNR) at the UAV is given by
	\begin{equation}
		{\rm SNR}_{\rms}(d,\theta)
		=
		\frac{
			\trp \, G_{\mathrm{eff}}^2({\Theta_{\ob}})\, \lambda_c^2 \, \sigma
		}{
			(4\pi)^3 \, d^{2\alpha} \, N_0
		}
		\exp\!\left(-2\kappa_f d\right),
	\end{equation}The factor $\lambda_c^2/(4\pi)^3$ is the two-way radar path-loss constant, where $d \triangleq d(r) = \sqrt{r^2+H^2}$ is the three-dimensional (slant) distance between the UAV at altitude~$H$ and a vehicle at horizontal distance~$r$, $\lambda_c={c}/{f_c}$ 
is the carrier wavelength ($c$ denotes speed of light, $f_c$ denotes the carrier frequency), $N_{0}$ is the additive white Gaussian noise (AWGN) power, 
and $G_{\mathrm{eff}}({\Theta_{\ob}})$ is 
the effective monostatic sensing gain defined in \eqref{eq:Geff}.
 
 {\begin{definition}[Detection probability]
 	A vehicle at $\x=(r,\theta)$ is successfully detected if $\mathrm{SNR}_{\rms}(d(r),\theta)\ge \tau_\rms$,
 	where $\tau_\rms$ is the sensing threshold. The DP of a vehicle at distance
 	$r$ is
 	\begin{align}
 		{\rm P_d}(r)=\mathbb{P}[\mathrm{SNR}_\rms(d(r),\theta)\ge \tau_\rms]. \label{Pdr}
 	\end{align}
 \end{definition}
\noindent The SNR-threshold detection model is standard in radar theory where a target is declared present when the received echo exceeds the noise floor by a factor~$\tau_{\rms}$. When multiple vehicles fall within the same beam sector simultaneously, each vehicle generates an independent echo that can be resolved in range (time-delay) by the radar waveform. Cross-target interference from nearby vehicles is negligible at millimeter-wave frequencies due to the narrow beamwidth. We therefore model detections as independent across vehicles, conditioned on their respective distances. {Since the UAV performs a full $2\pi$ angular scan, every vehicle within horizontal distance~$r_{\rms}$ is guaranteed to be illuminated by the beam during the scan cycle. Consequently, the angular alignment factor $\gamma/(2\pi)$ in the DP~\eqref{Pdr} is immaterial for determining the sensing range, the only relevant question is whether the SNR and LoS conditions are sufficient {when the beam points at the vehicle}. Therefore to define the maximum reliable sensing distance, we define the conditional DP as
	\begin{align}
		{\rm P_d^{cond}}(r)\triangleq p_{\sf L}(r,H)\,\exp\!\left(-\frac{\tau_s(4\pi)^3 d^{2\alpha} N_0\, e^{2\kappa_f d}}{P_t G_s^2(\gamma) \lambda_c^2 \bar\sigma}\right),\label{eq:Pd_cond}
	\end{align}
	where $d\triangleq d(r)=\sqrt{r^2+H^2}$.} To guarantee a minimum sensing reliability, we impose a conditional DP constraint and define the maximum reliable sensing radius as follows.
\begin{figure}[t!]
	\centering
	\includegraphics[width=0.7\linewidth]{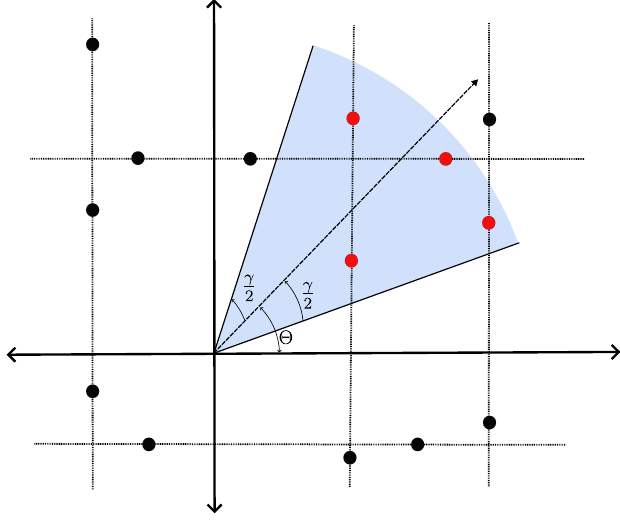}
	\caption{UAV-mounted RADAR sensing model with antenna orientation ${\Theta_{\ob}}$ and sensing sector bounded by ${\Theta_{\ob}}\pm\gamma/2$. Ground vehicles are distributed on MPLP roads with PPP-distributed vehicle locations on each road.}
	\label{fig:systemmodelantenna}
\end{figure}
\begin{definition}[Maximum reliable sensing distance]
	The maximum reliable horizontal sensing distance $r_\rms$ is the largest horizontal distance at which a ground vehicle can be reliably detected,
	\begin{align}
		r_{\rm s} = \sup\!\big\{r \geq 0 : {\rm P_d^{cond}}(r) \geq \delta\big\},\label{eq:rs_def}
	\end{align}
	where $\delta\in(0,1)$ is the reliability threshold. Since ${\rm P_d^{cond}}(r)$ decreases monotonically in $r$, the sensing radius is jointly constrained by the SNR and blockage limits,
	\begin{align}
		r_{\rm s} = \min\!\left(r_{\rm s}^{\rm snr},\; r_{\rm L}\right),\label{eq:rs_min}
	\end{align}
	where $r_{\rm s}^{\rm snr} = \sqrt{d_{\rm s}^2 - H^2}$ and $r_{\rm L}=H/\tan(\pi a/180)$.
\end{definition}
 \begin{definition}[UAV sensing zone]
For the typical UAV during a single scan step with boresight direction~${\Theta_{\ob}}$, the instantaneous sensing zone is defined as the set
 	\begin{equation}
 		\mathcal{S}({\Theta_{\ob}})
 		\triangleq
 		\left\{
 		\x\in\Psi :
 		r \le r_\rms,
 		\;
 		\left|\theta-{\Theta_{\ob}}\right| \le {\gamma}/{2}
 		\right\},
 		\label{eq:sensing_zone}
 	\end{equation}
 	where $\x=(r,\theta)$ denotes the location of a ground vehicle,
 	$r_\rms$ is the maximum reliable sensing distance,
 	${\Theta_{\ob}}$ is the {typical UAV's} antenna boresight direction,
 	and $\gamma$ is the antenna beamwidth. {For an arbitrary UAV~$\y_i$ the sensing zone is $\mathcal{S}(\Theta_i)$, with $\Theta_i$ being its independent mark.}
 \end{definition}{\noindent Since each UAV~$\y_i$ has an independent mark~$\Theta_i$,} the overall sensing region is modeled as a Boolean process given by
 \begin{equation}
 	\mathcal{B}
 	\triangleq
 	\bigcup\nolimits_{\y_i\in\Phi_u} \left( \y_i + \mathcal{S}_i({\Theta_i}) \right),
 \end{equation}which represents the spatial region sensed by all the UAVs during the sensing phase. Note that the sensing set $\mathcal{S}$ is convex, being the intersection of a disk
 and an angular sector. Hence, each translated set $\mathbf{y}_i+\mathcal{S}$
 is also convex. \subsection{Communication model}
{Within each sector, immediately after sensing, the UAV communicates with the detected vehicles using the beam alignment just obtained. If $L$ vehicles were detected in that sector, the UAV allocates $B/L$ of its bandwidth to each. Sectors with no detection are skipped. Because the beam illuminates one sector at a time, vehicles in different sectors do not compete for bandwidth simultaneously.} We now characterize the communication link model and the performance metrics.\\
 \noindent{\textit{Directional antenna model:}  Under the sector-based antenna model, the effective gain between
 	the typical UAV and the vehicle is 
 	\begin{align}
 		G_{\crm}(\theta)
 		=
 		G_\crm\,
 		\1 \!\left(
 		|\theta-{\Theta_{\ob}}|\le \gamma/2
 		\right),
 	\end{align}where $G_\crm$ denotes the main-lobe communication gain.  \\
 	\noindent\textit{SINR formulation:} 
 	By construction, the desired communication link is
 	always aligned with the main lobe once a vehicle is detected. Accordingly, the downlink ${\rm SINR}$ at the
 	considered vehicle is given by
 	\begin{align}
 		\SINR
 		=
 		\frac{
 			P_\trm\, G_\crm\, h_{\ob}\, d^{-\alpha}
 			e^{-\kappa_f d}
 		}{
 			\displaystyle
 			\sum\nolimits_{\y \in \Phi_u^{\rm a}\setminus \{\ob\}}
 			P_\trm\, G_{\y}\, h_{\y}\,
 			d_{\y}^{-\alpha}
 			e^{-\kappa_f d_{\y}}
 			+N_{0}
 		},
 	\end{align}
 	where  $h_{\ob}$ and $h_{\y}$
 	are independent Rayleigh fading gains and
 	$d_{\y}$ is the distance between the interfering UAV $\y$ and the vehicle. For an interfering UAV $\y$, the effective communication gain is given by
 	\begin{align}
 		G_{\y}
 		=
 		G_\crm\,
 		\1\!\left(
 		|\theta_{\y}-\Theta_{\y}|\le \gamma/2
 		\right),
 	\end{align}
 	where $\theta_{\y}$ denotes the angular position of the considered vehicle with
 	respect to UAV $\y$, and $\Theta_{\y}$ is the beam orientation of UAV $\y$. This captures the fact that interference is present only
 	when the interfering UAV’s beam is aligned with the vehicle. The CP is defined as the probability that the received SINR exceeds a target threshold $\tau_{\crm}$, i.e.,
 	\begin{align}
 		{\rm p_{c}}(\tau_{\crm}) = \mathbb{P}(\SINR > \tau_{\crm}).
 	\end{align}
\begin{remark}\label{rem:scope}
{Vehicles are modeled at fixed positions during a single {sector's sense-communicate interval}. This is reasonable because typical cycle durations are short enough that even fast-moving vehicles traverse a negligible fraction of the sector width, so beam alignment obtained during sensing remains valid for communication. Across successive {sectors and} cycles, vehicles may enter or leave the sensing region. Because the underlying point processes are stationary, each {sector} constitutes an independent spatial snapshot.} 
\end{remark}

\begin{definition}[Full-scan sensing region]
	The full-scan sensing region of a UAV is defined as the union of its instantaneous
	sensing zones over a complete angular scan, i.e.,
	\begin{equation}
		\mathcal{S}_{\rm f}\equiv \bt_{2}(\ob,r_{\rms})
		\triangleq
		\bigcup\nolimits_{{\Theta_{\ob}}\in[0,2\pi)} \mathcal{S}({\Theta_{\ob}}).
		\label{eq:scan_region}
	\end{equation}
\end{definition}
\begin{definition}[UAV activation probability]
	The UAV activation probability, denoted by ${\rm p_a}$, is defined as the probability
	that the typical UAV sense
 at least one vehicle within its its full scan of sensing zone. Precisely we can write it as
	\begin{equation}
		{\rm p_a}
		\triangleq
		\mathbb{P}\!\left(
		\Psi\big(\bt_{2}(\ob, r_\rms)\big) \ge 1
		\;\middle|\;
		\ob \in \Phi_u
		\right).
	\end{equation}
\end{definition}

\begin{figure}[t!]
	\centering
	\resizebox{\columnwidth}{!}{%
	\begin{tikzpicture}[
		timeline/.style={thick,->},
		block/.style={
			draw,
			thick,
			rounded corners=2pt,
			minimum height=0.7cm,
			minimum width=1.3cm,
			align=center,
			font=\scriptsize
		},
		font=\footnotesize
		]
		
		\draw[timeline] (0,0) -- (9.8,0) node[right]{Time};
		
		\node[block, fill=gray!15] (s1) at (1.0,0)
		{Sense\\[-1pt]Sec.~1};
		
		\node[block, fill=blue!15] (c1) at (2.6,0)
		{Comm\\[-1pt]Sec.~1};
		
		\node[block, fill=gray!15] (s2) at (4.2,0)
		{Sense\\[-1pt]Sec.~2};
		
		\node[block, fill=blue!15] (c2) at (5.8,0)
		{Comm\\[-1pt]Sec.~2};
		
		\node at (7.0,0) {$\cdots$};
		
		\node[block, fill=gray!15] (sN) at (8.2,0)
		{Sense\\[-1pt]Sec.~$N_s$};
		
		\draw[->, thick] (s1.east) -- (c1.west);
		\draw[->, thick] (c1.east) -- (s2.west);
		\draw[->, thick] (s2.east) -- (c2.west);
		
	\end{tikzpicture}}
	
	\caption{{Sense-then-communicate cycle of a UAV with $N_s=2\pi/\gamma$ sectors. In each sector, the UAV first senses and then communicates with detected vehicles before advancing to the next. If no vehicle is detected in any sector, the UAV remains idle for the entire cycle.}}
	\label{fig:scan_sensing_comm_timeline}
\end{figure}
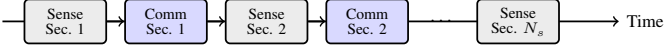
 \noindent Let $R_1$ and $R_k$ denote the distances from the typical UAV to its nearest and
	$k$th nearest detected vehicle in the sector respectively. The corresponding CP is defined as follows.
	
	\begin{definition}[Nearest and $k$th-nearest vehicle coverage]
		The nearest and $k$th-nearest vehicle coverage probabilities are defined as \begin{align}
			&	{\rm p}^{1}_{\rm c}
			=
			\mathbb{P}\!\left(
			\SINR_{1} > \tau_{\crm}
			\;\middle|\;
			R_{1}\le r_{\rms}
			\right)	\\&		{\rm p}^{k}_{\rm c}
			=
			\mathbb{P}\!\left(
			\SINR_{k} > \tau_{\crm}
			\;\middle|\;
			R_k \le r_{\rms}
			\right),
		\end{align} where $r_{\rms}$ denotes the sensing radius.
	\end{definition}
	
\noindent	The downlink rate of the $k$th served vehicle is given by
	\begin{align}
		\mathcal{R}
		=
		({B}/{L})\log_2\!\left(1+\SINR_k\right),\label{RC}
	\end{align}
	where $B$ is the system bandwidth and {$L$ denotes the number of vehicles detected by the UAV within the current angular sector of beamwidth~$\gamma$}. Note that as the sensing radius~$r_{\rm s}$ grows (e.g., via higher transmit power or narrower beamwidth), the UAV detects more vehicles on average, which increases~$L$ and reduces the per-vehicle rate~$B/L$. This creates a fundamental {sensing--rate trade-off}: better sensing coverage comes at the cost of lower per-user throughput. The corresponding RC is defined as
	\begin{align}
	{r_{\crm}}(\tau)
		=
		\mathbb{P}(\mathcal{R} > \tau \mid R_1 \le r_{\rms}),
	\end{align}{which conditions on the nearest vehicle falling in the sensing zone ($R_1\le r_{\rms}$, equivalently $N(r_{\rms})\ge 1$). More generally, the {$k$th-nearest RC} conditions on the $k$th nearest vehicle being sensed
	\begin{align}
	{r_{\crm}^{k}}(\tau)
		=
		\mathbb{P}(\mathcal{R} > \tau \mid R_k \le r_{\rms}).
	\end{align}
{To account for the possibility that fewer than $k$ vehicles exist in the sensing zone, we define the typical RC as
	\begin{align}
		\tilde{r}_{\crm}^{k}(\tau)
		=
		\mathbb{P}(R_k \le r_{\rms}) \cdot r_{\crm}^{k}(\tau),\label{eq:eff_rc}
	\end{align}
	where $\mathbb{P}(R_k \le r_{\rms})$ is the probability that at least $k$ vehicles lie within $r_{\rms}$. This product captures the system-level coverage performance}
	which jointly captures both sensing success and communication performance. For the nearest vehicle ($k=1$), $\mathbb{P}(R_1\le r_{\rms})=1-p_0(r_{\rms})={\rm p_a}$, so $\tilde{r}_{\crm}^{1}(\tau)={\rm p_a}\cdot r_{\crm}(\tau)$.}
}}

{\section{Sensing Analysis}
{We first derive the DP and the maximum reliable sensing distance. Without loss of generality, we assume that the typical UAV is located at the origin.}}  
 
{\begin{theorem} \label{Detection-prob}
		For a vehicle located at distance $r$ from the typical UAV, the blockage-aware detection
		probability under directional sensing is given by (for proof see Appendix \ref{proofthm1})
		\begin{align}
			{\rm P_{d}}(r)
			=\frac{\gamma}{2\pi}\; p_{\sf L}(r,H) \;\exp\!\left(-\frac{\tau_s(4\pi)^3 d^{2\alpha} N_0\, e^{2\kappa_f d}}{P_t G_s^2(\gamma) \lambda_c^2 \bar\sigma}\right),\label{detectionprobability}
		\end{align}where $d=\sqrt{r^2+H^2}$ and $p_{\sf L}(r,H)$ is the LoS probability given in \eqref{eq:PLoS}. The factor $p_{\sf L}(r,H)$ reflects the requirement that the radar round-trip must traverse a LoS path, vehicles in NLoS experience round-trip attenuation that renders negligible detection .
	\end{theorem}
}{\begin{remark}[Role of blockage in DP]
The DP in \eqref{detectionprobability} is the product of three terms: (i)~the angular alignment probability $\gamma/(2\pi)$, (ii)~the LoS probability $p_{\sf L}(r,H)$, and (iii)~the exponential SNR decay. Among these, the LoS probability introduces a distance-dependent ceiling on the DP that is independent of the transmit power. Due to the step-like behavior of $p_{\sf L}(r,H)$ described in~\eqref{eq:PLoS_step}, the DP exhibits a sharp cutoff at the blockage-limited distance~$r_{\rm L}$, beyond which detection becomes negligible regardless of the radar power budget. 
\end{remark}}
\noindent{{Using the conditional DP ${\rm P_d^{cond}}(r)$ from~\eqref{eq:Pd_cond}, we now derive the maximum reliable sensing distance.}}

{\begin{theorem}
 The maximum
	horizontal sensing radius $r_{\rm s}$ for a UAV at altitude $H$ is
\begin{align}
		r_{\rm s}
		=
		\min\!\left(\sup\!\big\{r \geq 0 : {\rm P_d^{cond}}(r) \geq \delta\big\},\; r_{\rm L}\right),
	\end{align}
where ${\rm P_d^{cond}}(r)$ is defined in~\eqref{eq:Pd_cond} and $r_{\rm L}=H/\tan(\pi a/180)$ is the blockage-limited radius from~\eqref{eq:PLoS_step}.
	The SNR-limited radius $r_{\rm s}^{\rm snr}=\sqrt{d_{\rm s}^2-H^2}$ is the power-limited radius with $d_{\rm s}$ satisfying
	\begin{align}
		d_{\rms}
		=
		\frac{\alpha}{\kappa_f}
		\,{\rm W}\!\left(
		\frac{\kappa_f}{\alpha}
		\left(
		\frac{-\ln(\delta)\, \trp G_s^2(\gamma) \lambda_c^2}
		{\tau_{\rms} (4\pi)^3 N_0}
		\right)^{\frac{1}{2\alpha}}
		\right),
	\end{align}and consequently $r_{\rm s}=\min(r_{\rm s}^{\rm snr},\, r_{\rm L})$. (For proof see Appendix \ref{proofthm2}.)
\end{theorem}}
{\begin{remark}
The sensing radius $r_{\rm s} = \min(r_{\rm s}^{\rm snr}, r_{\rm L})$ identifies two regimes. In the \textit{blockage-dominant regime} ($r_{\rm L} < r_{\rm s}^{\rm snr}$), urban obstructions limit sensing range before SNR becomes insufficient, which is common at low altitudes in dense areas where increasing transmit power yields no gain. In the \textit{SNR-dominant regime} ($r_{\rm s}^{\rm snr} < r_{\rm L}$), path loss and molecular absorption are the bottleneck, occurring at higher altitudes with large elevation angles. The transition is governed by altitude $H$ and environment parameters $(a,b)$. In the blockage-dominant regime altitude should be increased to widen the LoS cone whereas in the SNR-dominant regime altitude should be minimized to reduce path loss.
\end{remark}}

\noindent{We now derive the distribution of the number of detected vehicles within {a single angular sector of beamwidth~$\gamma$}. This distribution is a key building block to determine the per-sector  communication load, and ultimately the RC (since the per-vehicle rate is $B/L$).} We first present the result for beamwidth $0<\gamma<\pi$, we then present a generalize result when $\gamma=2\pi$ that is the UAV equipped with an omnidirectional antenna. 
{\begin{theorem}
		The probability generating function (PGF) of the number of detectable vehicles $N(r_{\rms})$
		falling within the sector sensing region of radius $r_{\rms}$ around the typical UAV
		is given by (for proof sec Appendix \ref{proofthmpmf})
		\begin{align}
			\PGF_{N(r_{\rms})}(s)
			&=
			\exp\!\big( g_\gamma(s,r_{\rms}) \big),
		\end{align}
		where
\begin{align}
	&g_\gamma(s,r_{\rms})=
	\lambda_{\ell}
	\left[2
	\int_{0}^{r_{\rm s}\sin(\gamma/2)}
\left(
\exp\!\left(\Lambda_\rmh(\rho)(s-1)\right)-1\right)\,
	\dv \rho\right.
	\nonumber\\
	&\left.\qquad
	+
	\int_{0}^{r_{\rm s}}
\left(
\exp\!\left(\Lambda_\rmv(\rho)(s-1)\right)-1\right)\,\dv \rho
	\right],
\end{align}and the average number of vehicle on each chord are
	\begin{align}
	&\Lambda_\rmh(\rho)
	=
	\int_{\ell_\rmh(\rho,\gamma)} \lambda\; p_{\sf L}\!\big(\!\sqrt{x^2+\rho^2},\,H\big)\;\dv x, \label{Lambdah} \\
	&\Lambda_\rmv(\rho)
	=
	\int_{\ell_\rmv(\rho,\gamma)} \lambda\; p_{\sf L}\!\big(\!\sqrt{\rho^2+y^2},\,H\big)\;\dv y.\label{Lambdav}
\end{align}The chord limits $\ell_\rmh(\rho,\gamma)$ and $\ell_\rmv(\rho,\gamma)$ have lengths
	\begin{align}
	&|\ell_\rmv(\rho,\gamma)|
	=
	2\,\min\!\left(
	\sqrt{r_s^{2}-\rho^{2}},
	\;
	\rho\,\tan\!\left({\gamma}/{2}\right)
	\right), \label{vchord} \\
	&|\ell_\rmh(\rho,\gamma)|
	=
	\sqrt{r_{\rm s}^{2}-\rho^{2}}
	-
	\rho\,\cot \!\left({\gamma}/{2}\right).\label{hcord}
\end{align}When $p_{\sf L}\equiv 1$ (no blockage), $\Lambda_\rmh(\rho)$ reduces to $\lambda|\ell_\rmh(\rho,\gamma)|$, recovering the original expressions.

	\end{theorem}
	}

\noindent Using the PGF, we can derive the PMF and perform the statistical analysis of the sensing load $N({r_\rms})$
\begin{corollary}\label{cor:PMFloadtypicalcell}
	The PMF $\mathbb{P}\left({N}(r_{\rms})=k\right)$ of the sensing load ${N}(r_{\rm s})$ on the typical UAV
	\begin{align}
		&p_{k}(r_{\rms})=\mathbb{P}\left({N(r_{\rms})}=k\right)\notag\\
		&=e^{f_{0}(r_{\rms})}\sum\nolimits_{\mathrm{N}_{k}}\frac{\left(f_{1}(r_{\rms})\right)^{n_{1}}\cdots\left(f_{m}(r_{\rms})\right)^{n_{m}}}{n_{1}!\cdots n_{m}!},\label{PMF}\\
		&\text{where }f_{0}(r_{\rms})={g(0,r_{\rms})},
		\quad f_{m}(r_{\rms})={g^{(m)}(0,r_{\rms})}/{m!},\text{ with}\notag\\
		& g^{(m)}(0,r_{\rms})=
		\lambda_{\ell}\left[
		2\int_{0}^{r_{\rm s}\sin(\gamma/2)}
		\big(\Lambda_{\rm h}(\rho)\big)^m
		e^{-\Lambda_{\rm h}(\rho)}\,\mathrm{d}\rho\right.
		\nonumber\\
		&\left.+\int_{0}^{r_{\rm s}}
		\big(\Lambda_{\rm v}(\rho)\big)^m
		e^{-\Lambda_{\rm v}(\rho)}\,\mathrm{d}\rho
		\right],
	\end{align}where $\Lambda_{\rm h}(\rho)$ and $\Lambda_{\rm v}(\rho)$ are the LoS-thinned chord counts from \eqref{Lambdah}--\eqref{Lambdav}.
\end{corollary}

%

\begin{corollary}
	The CDF of the distance of the $n$th nearest vehicle from the UAV is $F_{R_{n}}(r)=$
	
	\begin{align}
		1-\sum\nolimits_{k=0}^{k-1}e^{f_{0}(r)}\sum\nolimits_{\mathrm{N}_{k}}\frac{\left(f_{1}(r)\right)^{n_{1}}\cdots\left(f_{m}(r)\right)^{n_{m}}}{n_{1}!\cdots n_{m}!}\label{CDFRn}
	\end{align}
	The PDF of the distance $R_n$ of the $n$th
	nearest vehicle from the UAV is given by
	\begin{align}
		f_{R_n}(r)
		=
		-\sum\nolimits_{k=0}^{n-1}\frac{\dv}{\dv r}p_k(r).
		\label{PDFRn}
	\end{align}
\end{corollary}\noindent Using the CDF of the $n$th nearest vehicle, we now present the CDF and PDF 
for the nearest vehicle from the UAV in the following corollary.
{\begin{corollary}
		For the nearest detectable vehicle $R_1$, the CDF and PDF are given by
		\begin{align}
			F_{R_{1}}(r) = 1 - e^{f_{0}(r)},\,\,
			f_{R_{1}}(r) = -f^{(1)}_{0}(r)e^{f_{0}(r)},\\
			\text{where }f_{0}(r)
			=
			\lambda_{\ell}
			\Bigg[
			2
			\int_{0}^{r\sin(\gamma/2)}
			\left(
			e^{-\Lambda_{\rm h}(\rho)}
			-1
			\right)
			\mathrm{d}\rho
			\nonumber\\
			\qquad +
			\int_{0}^{r}
			\left(
			e^{-\Lambda_{\rm v}(\rho)}
			-1
			\right)
			\mathrm{d}\rho
			\Bigg],
		\end{align}
		and $f^{(1)}_{0}(r) = \frac{\dv}{\dv r}f_0(r)$, where $\Lambda_{\rm h}(\rho)$ and $\Lambda_{\rm v}(\rho)$ are the LoS-thinned chord counts from \eqref{Lambdah}--\eqref{Lambdav} evaluated with the sensing sector of radius $r$.
	\end{corollary}

\noindent The following corollaries translate the distance distributions derived above into operationally relevant sensing metrics for the typical UAV. 	Let the distance of the nearest and $n$th nearest vehicle from the typical UAV is denoted as $R_{1}$ and $R_{n}$. This can be directly obtained using the CDF of $R_{n}$ presented in \eqref{CDFRn}.

\begin{corollary}
	The probability that the nearest and $k$th nearest vehicle falls under the sensing zone of the typical UAV is
	\begin{align}
		\mathbb{P}\left(R_{1}\leq r_{\rm s}\right)=F_{R_{1}}(r_{\rm s}),\,\,
		\mathbb{P}\left(R_{n}\leq r_{\rm s}\right)=F_{R_{n}}(r_{\rm s})
	\end{align}
\end{corollary}
\begin{corollary}\label{thm:Ntilde}
The PGF of the number of detectable vehicles located within the UAV sensing region
$\bt_{2}(\mathbf{o}, r_{\rm s})$ for $\gamma = 2\pi$, corresponding to an omnidirectional antenna. 
	\begin{align}
		&\PGF_{{N}(r_{\rms})}(s)=\expS{g_{2\pi}(s,r_{\rms})},\\
		&\text{where}\quad	g_{2\pi}(s,r_{\rms})=	4\lambda_{\ell}\left(\int_{0}^{r_{\rms}}e^{\Lambda(\rho)(s-1)}\dv \rho-r_{\rms}\right),\nonumber\\
		&\text{with}\quad \Lambda(\rho)= 2\int_{0}^{\sqrt{r_{\rms}^{2}-\rho^{2}}}\lambda\, p_{\sf L}\!\big(\!\sqrt{x^2+\rho^2},H\big)\dv x.\nonumber
	\end{align}
\end{corollary}


\begin{corollary}
	The probability that no vehicle is present within the sensing region
	\begin{align*}
		&{\rm p}_{\rm idle}=\mathbb{P}\left[{N}(r_{\rms})=0\right]=\expS{4\lambda_{\ell}\left(\int_{0}^{r_{\rms}}e^{-\Lambda(\rho)}\dv \rho-r_{\rms}\right)},
	\end{align*}where $\Lambda(\rho)= 2\int_{0}^{\sqrt{r_{\rms}^{2}-\rho^{2}}}\lambda\, p_{\sf L}(\sqrt{x^2+\rho^2},H)\dv x$.
\end{corollary}
\begin{table}[t!]
	\caption{System parameters.}
	\label{tab:detection_params}
	\centering
	\small
	\begin{tabular}{ll}
		\hline
		\textbf{Parameter} & \textbf{Value} \\
		\hline
		Vehicle intensity, $\lambda$ & $30~\si{vehicles/km}$ \\
		Road density, $\lambda_l$ & $2~\si{roads/km}$ \\
		UAV altitude, $H$ & $100~\si{m}$ \\
		UAV density, $\lambda_u$ & $5~\si{UAV/km^2}$ \\
		Transmit power, $P_t$ & $1$ (normalized) \\
		{Sensing/comm.\ gain, $G_s(\gamma){=}G_c(\gamma)$} & {$2\pi/(3\gamma)$} \\
		Carrier wavelength, $\lambda_{\rm c}$ & $0.005~\si{m}$ \\
		Noise power, $N_0$ & $10^{-9}$ \\
		Sensing SNR threshold, $\tau_{\rm s}$ & $1$ \\
		Mean fading power, $\bar{\sigma}$ & $1$ \\
		{Absorption coefficient, $\kappa_f$} & {$3.45~\text{Np/km}$~\cite{ITU-R-P676}} \\
		Path-loss exponent, $\alpha$ & $2.2$ \\
		\hline
		\multicolumn{2}{l}{\textbf{LoS parameters} ($a$, $b$):} \\
		\quad Suburban       & $4.88,\ 0.43$ \\
		\quad Urban          & $12.08,\ 0.11$ \\
		\quad Dense Urban    & $9.61,\ 0.16$ \\
		\quad High-rise Urban & $27.23,\ 0.08$ \\
		\hline
	\end{tabular}
\end{table}
\noindent Hence the UAV activation probability is ${\rm p_{a}}=1-{\rm p}_{\rm idle}$. Observe that ${\rm p}_{\rm idle}$ is governed primarily by whether any road from the MPLP intersects the UAV's sensing disk of radius~$r_{\rm s}$: if no road passes through the disk, no vehicle can be present regardless of the vehicular density~$\lambda$. Consequently, the activation probability depends on the UAV's position relative to the road layout, captured through the line density~$\lambda_{\ell}$ and the sensing radius~$r_{\rm s}$.
\begin{remark}
	As $\lambda \to \infty$, the integral term vanishes and the idle probability converges to $\exp(-4\lambda_{\ell} r_s)$. Consequently, the UAV activation probability saturates at
	\begin{align}
		\lim\nolimits_{\lambda \to \infty} {\rm p_{a}}
		= 1 - \exp\!\left(-4\lambda_{\ell} r_s\right).\label{asym}
	\end{align}
	{This limit equals the probability that at least one road from the MPLP intersects the UAV's sensing disk of radius~$r_{\rm s}$. When vehicles are abundant, every road that crosses the disk contains at least one vehicle, hence the bottleneck is whether any road is present at all.}
\end{remark}\noindent Having characterized the sensing metrics---DP, sensing radius, vehicle count distribution, and activation probability---we now analyze the communication performance of the detected vehicles. 
\section{Communication Analysis}
{{In the communication phase of each sector, the active UAV steers its directional beam using the angular location obtained from the immediately preceding sensing step. A detected vehicle is served by its nearest active UAV, all other active UAVs whose beams happen to illuminate the vehicle contribute to interference.}}  Hence we now present the CP for the nearest and $k$th nearest vehicle from the typical UAV. Let the nearest and the $k$th nearest vehicle is located at the $\x_{1}$ and $\x_{k}$. Not that the nearest and the $k$th nearest vehicle may be located on the same or the different road. For the $k$-th nearest vehicle located at $\x_{k}$, the aggregate LoS-only interference is given by
\begin{align}
	I_k&=\sum_{\y  \in \Phi_u^{\rm a}\setminus \bt_{2}(\x_k,r_k)}
	P_\trm\, G_{\y}\, h_{\y}\,
	d(\|\y-\x_{k}\|)^{-\alpha}\notag\\
	&\expS{-\kappa_f d(\|\y-\x_{k}\|)}\,\1(\text{LoS}),
\end{align}where $\1 (\text{LoS})$ is a Bernoulli indicator with probability $p_{\sf L}(\|\y-\x_k\|,H)$, reflecting that only LoS interferers contribute meaningfully.
{\begin{theorem}
 The LT $\mathcal{L}_{I_k}(s)$ of $I_k$ under the LoS-only interference model is (for proof see Appendix \ref{proofLT}) 
		\begin{align}
&=\exp\!\left(
			- {\gamma \lambda_u {\rm p_a}}
			\int_{r_k}^{\infty}\frac{{
				s P_{\trm} G_{\y} (d(x))^{-\alpha}
				e^{-\kappa_f d(x)}
			} \,p_{\sf L}(x,H)\, x \,\dv x}{
					1+
					s P_{\trm} G_{\y} (d(x))^{-\alpha}
					e^{-\kappa_f d(x)}
			}
			\right).\notag
		\end{align}The factor $p_{\sf L}(x,H)$ inside the integral captures the fact that only LoS interferers are considered.
	\end{theorem}}
{\begin{remark}
The effective density of interfering UAVs is $\lambda_u\,{\rm p_a}\,\gamma/(2\pi)$, arising from two successive independent thinnings of the marked PPP $\tilde{\Phi}_u=\{(\y_i,\Theta_i)\}$. First, a UAV is active if it detected at least one vehicle, yielding an independent ${\rm p_a}$-thinning that produces the active sub-process $\Phi_u^{\rm a}\sim\mathrm{PPP}(\lambda_u\,{\rm p_a})$. Second, by the independent marking theorem \cite{baccelli1997stochastic}, the beam orientation $\Theta_i$ is uniform on $[0,2\pi)$ and independent across UAVs due to MPLP isotropy, so the probability that a UAV's beam covers the typical vehicle is $\gamma/(2\pi)$. The product ${\rm p_a}\cdot\gamma/(2\pi)$ gives the effective interferer density in the LT. 
\end{remark}}
\noindent Equipped with the LT of interference and the distance distribution, 
we now present the CP for the $k$th nearest vehicle from the UAV.

	\begin{theorem}
		Conditioned on $R_k \le r_{\rms}$, the CP of the $k$-th nearest
		vehicle is given by (for proof see Appendix \ref{proof})
		\begin{align}
			&\prm^{k}_{\crm}(\tau_{\crm})=\frac{1}{\mathbb{P}(R_k \le r_{\rms})}\int_{0}^{r_{\rms}}
			\exp\!\left(
			-\frac{\tau_{\crm} N_{0} (d(r))^{\alpha} e^{\kappa_f d(r)}}{P_\trm G_\crm}
			\right)\notag\\
			&\mathcal{L}_{I_k}\left(\tau_{\crm} d^{\alpha}e^{\kappa_f d}
			\right)
			f_{R_k}(r)\dv r.\label{CP}
		\end{align}
	\end{theorem}

Now using the sensing load distribution, we present the rate CP in the following theorem.	
{\begin{theorem}
 		The rate CP of {the nearest vehicle served by the typical UAV},
 		conditioned on the event that at least one vehicle is present in the
 		sensing region, is given by (for proof see Appendix \ref{rcproof})
 		\begin{align}
 			r_{\crm}(\tau)&=
 			\frac{1}{1-p_0(r_{\mathrm{s}})}
 			\sum\nolimits_{k=1}^{\infty}
 			p_k(r_{\mathrm{s}})
 			\, {\rm p}_\crm^{k}\!\left(2^{\frac{\tau k}{B}}-1\right).
 		\end{align}
 	\end{theorem}
{\begin{corollary}
The RC of the $k$th nearest vehicle, conditioned on $R_k\le r_{\rms}$ (i.e., at least $k$ vehicles are detected in the sector), is
\begin{align}
	r_{\crm}^{k}(\tau)
	&=
	\frac{1}{\displaystyle\sum\nolimits_{j=k}^{\infty}p_j(r_{\mathrm{s}})}
	\sum\nolimits_{j=k}^{\infty}
	p_j(r_{\mathrm{s}})
	\,{\rm p}_\crm^{j}\!\left(2^{\frac{\tau j}{B}}-1\right).
\end{align}
For $k=1$ this reduces to the nearest-vehicle RC in Theorem~5, since $\sum_{j=1}^{\infty}p_j(r_{\mathrm{s}})=1-p_0(r_{\mathrm{s}})$.
\end{corollary}
\begin{corollary}
The typical RC is
\begin{align}
	&\tilde{r}_{\crm}^{k}(\tau)
	=
	\mathbb{P}(R_k\le r_{\rms})\cdot r_{\crm}^{k}(\tau)
	=
	\sum\nolimits_{j=k}^{\infty}
	p_j(r_{\mathrm{s}})
	\,{\rm p}_\crm^{j}\!\left(2^{\frac{\tau j}{B}}-1\right).\label{eq:eff_rc_k}\notag
\end{align}
For the nearest vehicle ($k=1$), $\tilde{r}_{\crm}^{1}(\tau)={\rm p_a}\cdot r_{\crm}(\tau)$. 
\end{corollary}}
 \noindent The typical RC reveals the sensing--rate trade-off: increasing $r_{\rms}$ raises the sensing probability $\mathbb{P}(R_k\le r_{\rms})$ but also increases the per-sector load~$L$, reducing the per-vehicle rate~$B/L$.  Consequently, the typical RC can be non-monotonic in the sensing radius.

%
%
 }

\section{Numerical Results}
We now present the analysis of the key analytical results, together with their Monte Carlo simulations. Unless stated otherwise, all parameter values are taken from Table~\ref{tab:detection_params}.

\begin{figure}
	\centering
	\includegraphics[width=\linewidth]{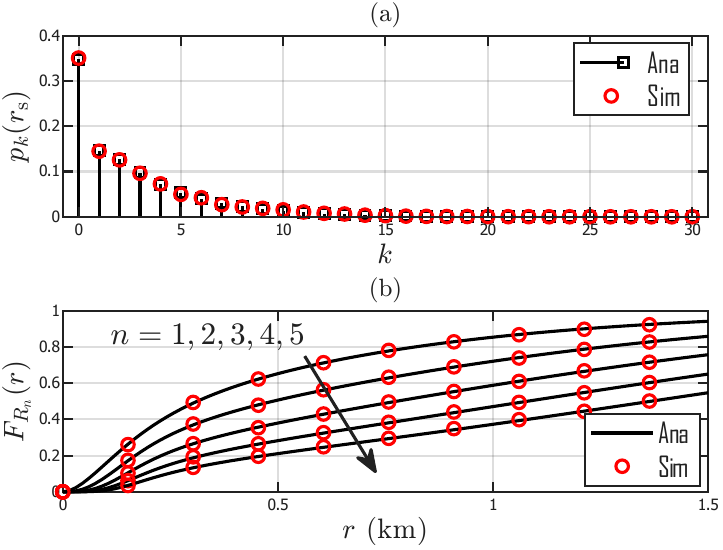}
	\caption{Analytical and simulation results 
	for the sensing load and the CDF of the $n$th nearest 
	{detectable} vehicle from the typical UAV. 
	(a)~PMF of the sensing load $p_k(r_{\rm s})$. 
	(b)~CDF of the distance $R_n$ to the $n$th nearest 
	{LoS-detectable} vehicle for $n=1,\ldots,5$.
	{The sensing radius is fixed at 
		$r_{\rm s} = 500~\si{m}$ with beamwidth $\gamma = \pi/6$, 
		UAV altitude $H = 100~\si{m}$, and LoS blockage parameters 
		$a = 9.61$, $b = 0.16$ (urban).}}
	\label{fig:pmfcdfblockageunified}
\end{figure}


\begin{figure}[t!]
	\centering
	\includegraphics[width=\linewidth]{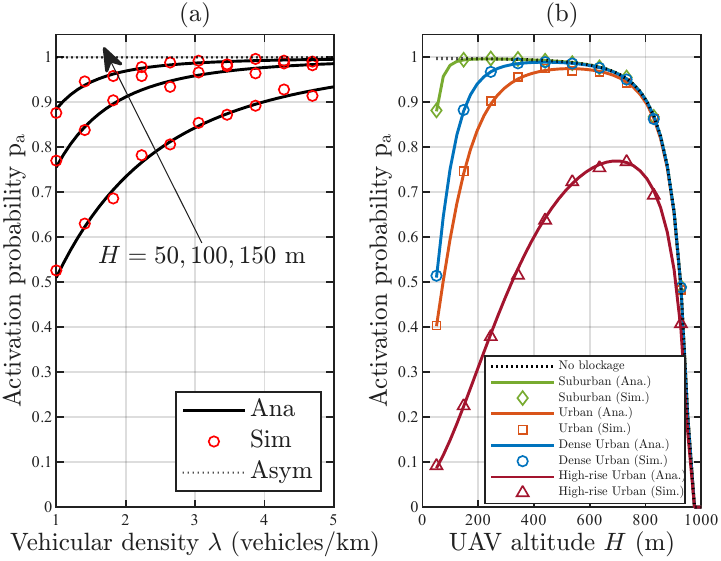}
	\caption{ (a) Activation probability ${\rm p_a}$ versus vehicular density $\lambda_v$ for $H \in \{50, 100, 150\}$~\si{m} in dense urban environment ($a = 9.61$, $b = 0.16$), with dashed lines indicating the asymptotic limit as $\lambda \to \infty$. (b) ${\rm p_a}$ with UAV altitude $H$ at $\lambda = 1$~vehicle/km, compared against a no-blockage.}
	\label{fig:activeprobwithh}
\end{figure}
\begin{figure}[t!]
	\centering
	\subfloat[]{\includegraphics[width=0.5\linewidth]{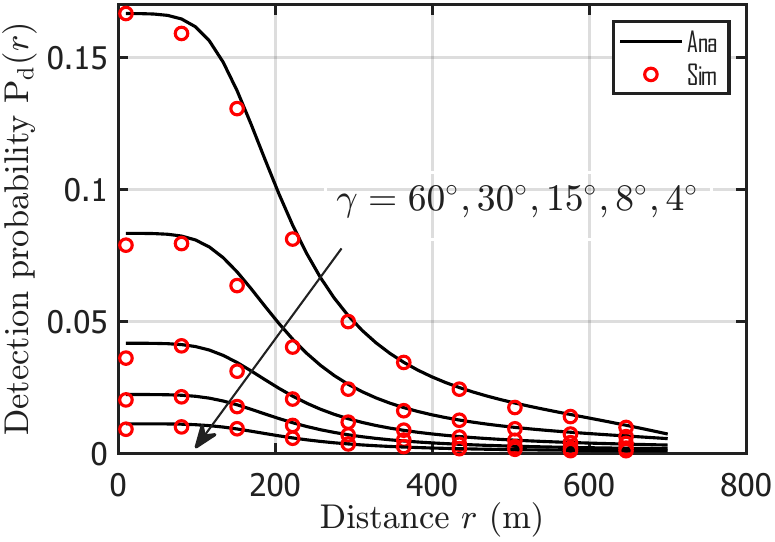}}%
	\hfil
	\subfloat[]{\includegraphics[width=0.5\linewidth]{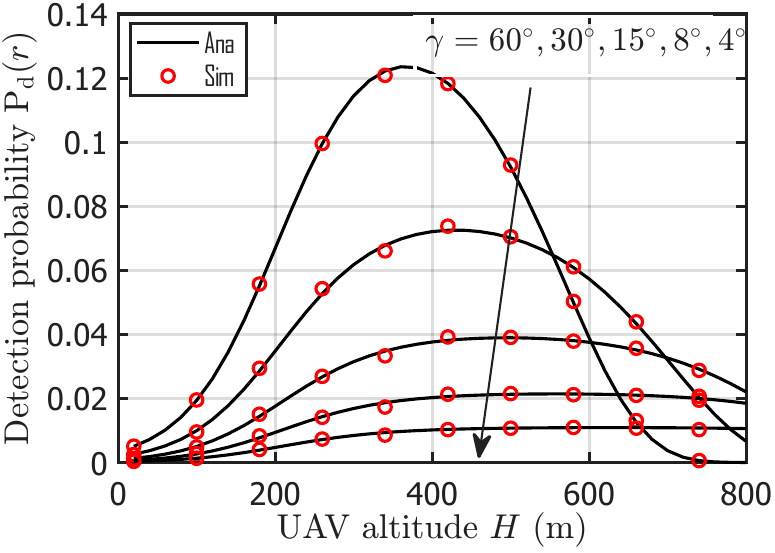}%
		\label{fig:pd_vs_H}}
	\caption{(a) DP ${\rm P_d}(r)$ under dense urban blockage ($a = 9.61$, $b = 0.16$) versus ground distance $r$ at fixed altitude $H = 100$~\si{m}; (b)${\rm P_d}(r)$ versus UAV altitude $H$ at fixed ground distance $r = 500$~\si{m}.}
	\label{fig:detection_prob}
\end{figure}
\subsection{Verification of analytical results}
To verify the {blockage-aware} distributions of the sensing 
load and the nearest vehicle distance, we plot the analytical and 
simulation results in Fig.~\ref{fig:pmfcdfblockageunified}. 
Fig.~\ref{fig:pmfcdfblockageunified}(a) shows the PMF of the sensing 
load given in~\eqref{PMF}. 
Fig.~\ref{fig:pmfcdfblockageunified}(b) shows the CDF of the distance 
to the $n$th nearest {LoS-detectable} vehicle from the typical 
UAV for $n = 1,\ldots,5$. The analytical and simulation results are in 
close agreement, validating the  analytical 
framework in the presence of blockage. The high value of $p_{\rm idle} = p_0(r_{\rms})$ in the PMF, 
despite the high vehicle density $\lambda = 30~\text{vehicles/km}$, is 
due to both the narrow beamwidth $\gamma = \pi/6$ {and the 
	LoS blockage at $H = 100~\si{m}$. 
}

\subsection{Impact of altitude on activation probability} {Fig. \ref{fig:activeprobwithh}(a) illustrates ${\rm p_a}$ versus the vehicular density $\lambda$ for three UAV altitudes in a dense urban environment. As expected, ${\rm p_a}$ increases monotonically with $\lambda$, since denser traffic raises the probability of at least one LoS-visible vehicle lying within the sensing region. For each altitude, ${\rm p_a}$ saturates toward the asymptotic limit $1 - \exp(-4\lambda_\ell r_\rms)$ as $\lambda \to \infty$, which depends only on the road density and effective sensing radius. At a fixed $\lambda$, ${\rm p_a}$ increases with altitude because a higher altitude steepens the elevation angle, improving $p_{\sf L}(r, H)$ and thereby increasing the number of sensed vehicles within the sensing region. Fig. \ref{fig:activeprobwithh}(b) plots ${\rm p_a}$ as a function of UAV altitude $H$ at $\lambda = 1$~\si{vehicle/km}. In the absence of blockage, ${\rm p_a}$ decreases monotonically with $H$, the growing slant distance amplifies path loss and molecular absorption, shrinking $r_\rms^{\text{snr}}$. All four environment curves first rise, peak, and then decline. The initial rise occurs because a steeper elevation angle improves $p_{\sf L}(r, H)$, increasing the effective density of sensed vehicle. Beyond a certain altitude, path loss and absorption dominate, and ${\rm p_{a}}$ falls. This non-monotonic behavior reveals a key design trade-off, the UAV altitude must balance LoS visibility gains against sensing range losses, yielding an environment-dependent optimal altitude that maximizes ${\rm p_a}$.}
\subsection{Analysis of the DP}
{Fig.~\ref{fig:detection_prob}(a) plots the  analytical and simulation values of DP ${\rm P_d}(r)$ with horizontal distance~$r$ for different  beamwidths at a fixed altitude $H = 100$~\si{m}.  For every beamwidth, ${\rm P_d}(r)$ decays with~$r$ due to increasing path loss, molecular absorption, and diminishing LoS probability. At a given distance, a wider beam yields a higher DP. This ordering arises because the angular alignment factor $\gamma/2\pi$ dominates the SNR gain. Fig.~\ref{fig:detection_prob}(b) shows ${\rm P_d}(r)$ with the UAV altitude~$H$ at a fixed ground distance $r = 500$~\si{m}. The initial rise is driven by improving LoS visibility as the elevation angle steepens with altitude, which increases the fraction of vehicles that are not blocked. Beyond a certain altitude, the growing slant distance amplifies path loss and molecular absorption, eventually dominating the LoS gain and causing ${\rm P_d}(r)$ to fall. The optimal altitude that maximises ${\rm P_d}(r)$ depends on the beamwidth. The widest beam ($\gamma = 60^\circ$) peaks earliest, because its moderate gain cannot sustain the SNR at large slant ranges. In contrast, the narrowest beam peaks much later,  since its higher gain extends the range over which sensing remains viable, allowing it to continue benefiting from improved LoS at greater altitudes. This confirms that directionality simultaneously governs both the sensing radius, through the gain-dependent SNR, and the angular coverage, through the beam-alignment probability, establishing a fundamental trade-off whose resolution depends jointly on the target range and deployment altitude.}
\subsection{Sensing radius under gain--beamwidth coupling}
Fig.~\ref{fig:sensingdesignchart2}(a) plots the maximum sensing radius $r_{\rms}$ as a function of the UAV altitude~$H$ for three detection thresholds with a fixed beamwidth $\gamma=\pi/6$. For each altitude, $r_{\rms}$ is computed as $\min(r_{\rm s}^{\rm snr},\,r_{\rm L})$, where
	$r_{\rm s}^{\rm snr}=\sqrt{d_{\rms}^2-H^2}$ is the power-limited
	horizontal radius obtained from Theorem~2 and
	$r_{\rm L}=H/\tan(\pi a/180)$ is the blockage-limited radius from the
	step-function approximation in~\eqref{eq:PLoS_step}. The initial rise is driven by the improving LoS probability as
	the elevation angle steepens with altitude. At low altitudes, the
	blockage-limited radius $r_{\rm L}$ is small because the shallow
	elevation angle results in frequent LoS obstruction by buildings, as $H$
	increases, $r_{\rm L}$ grows linearly, widening the effective sensing
	footprint. Beyond a certain $H$, the growing slant distance
	$d=\sqrt{r^2+H^2}$ amplifies both path loss ($d^{2\alpha}$) and
	molecular absorption ($e^{2\kappa_f d}$), causing $r_{\rm s}^{\rm snr}$
	to shrink faster than $r_{\rm L}$ grows, and $r_{\rms}$ decreases. Fig.~\ref{fig:sensingdesignchart2}(b) decomposes the sensing radius
	into its two  limits for $\delta=0.5$, the SNR-limited
	radius $r_{\rm s}^{\rm snr}$  decreases monotonically with
	$H$ as the slant distance grows, while the blockage-limited radius
	$r_{\rm L}$  increases linearly with~$H$ since a higher
	altitude widens the LoS cone. Their minimum reveals
	two distinct operating regimes. At low altitudes,
	$r_{\rm L}<r_{\rm s}^{\rm snr}$, placing the system in the {blockage-dominant regime} where the urban environment limits the
	sensing range regardless of the available transmit power. At higher
	altitudes, $r_{\rm s}^{\rm snr}<r_{\rm L}$, and the
	system enters the {SNR-dominant regime} where signal attenuation
	becomes the binding constraint. The crossover between these two regimes
	occurs at the altitude where $r_{\rm s}^{\rm snr}=r_{\rm L}$. This decomposition confirms
	that the optimal deployment altitude precisely balances LoS
	visibility gains against sensing range losses, and demonstrates that in
	the blockage-dominant regime, increasing transmit power or antenna gain
	yields no improvement in sensing coverage.

\begin{figure}
	\centering
	\includegraphics[width=\linewidth]{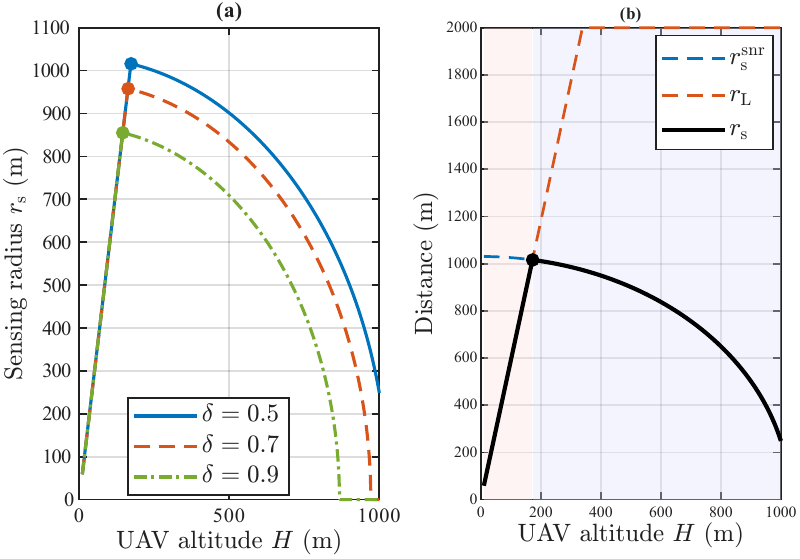}
	\caption{Sensing radius analysis with $\gamma=\pi/6$ in a dense urban	environment.
		(a)~Maximum sensing radius $r_{\rms}$ versus UAV altitude~$H$ for different.
		(b)~Decomposition of $r_{\rms}$ into the SNR-limited radius
		$r_{\rm s}^{\rm snr}$ and the blockage-limited radius
		$r_{\rm L}$  for $\delta=0.5$.}
	\label{fig:sensingdesignchart2}
\end{figure}
\subsection{Interference and CP analysis:}
Fig.~\ref{fig:LT_interference} presents the LT of interference at the $k$th nearest 
vehicle located at $R_k=100$~\si{m}. As UAV altitude increases, the LoS probability 
improves over a wider area, raising the activation probability~${\rm p_a}$ and 
consequently the density of active interferers, resulting in higher resultant 
interference and a lower LT. Fig.~\ref{fig:coverage_3gamma} shows 
that narrowing the beamwidth from $\gamma = \pi/3$ to $\gamma = \pi/12$ improves 
CP across the entire SINR threshold range. This is driven by two mechanisms: the 
antenna gain $G_\rms = 2\pi/(3\gamma)$ scales inversely with $\gamma$, strengthening 
the desired signal, while the angular fraction of interferers illuminating the 
typical UAV shrinks proportionally, reducing aggregate interference by a factor 
of $\gamma/(2\pi)$. Although a narrower beam extends 
the sensing radius $r_s$ via higher $G_{\rm s}$, thereby increasing ${\rm p_a}$, 
this effect is outweighed by the interference reduction, yielding a net 
improvement in CP.

\noindent Fig.~\ref{fig:covprobwithh}(a) shows the coverage probability (CP) of the nearest 
sensed vehicle versus UAV altitude $H$ for different UAV densities $\lambda_u$. 
The CP decreases monotonically with altitude as signal strength decreases with 
increasing path loss. At low altitudes, CP across different densities remains nearly 
identical, but the gap widens as $H$ increases. Although higher altitude reduces 
the activation probability and hence aggregate interference, the signal power 
degrades more severely, resulting in a net reduction in coverage. Higher UAV 
density further compounds this degradation through increased aggregate interference. 
Overall, altitude has a more dominant effect on coverage than UAV density, 
particularly at large $H$.} In Fig.~\ref{fig:covprobwithh}(b), CP is compared across 
different propagation environments. At low altitudes, all environments achieve 
near-identical coverage. However, as altitude increases, suburban environments 
suffer the most severe degradation due to the rapid decay of signal strength 
with distance in open areas. In contrast, high-rise urban environments benefit 
from building-induced blockage of interfering links, which preserves the 
signal-to-interference ratio and results in a significantly slower degradation 
of coverage with altitude.

\begin{figure}[t]
	\centering
	\subfloat[]{\includegraphics[width=0.45\columnwidth]{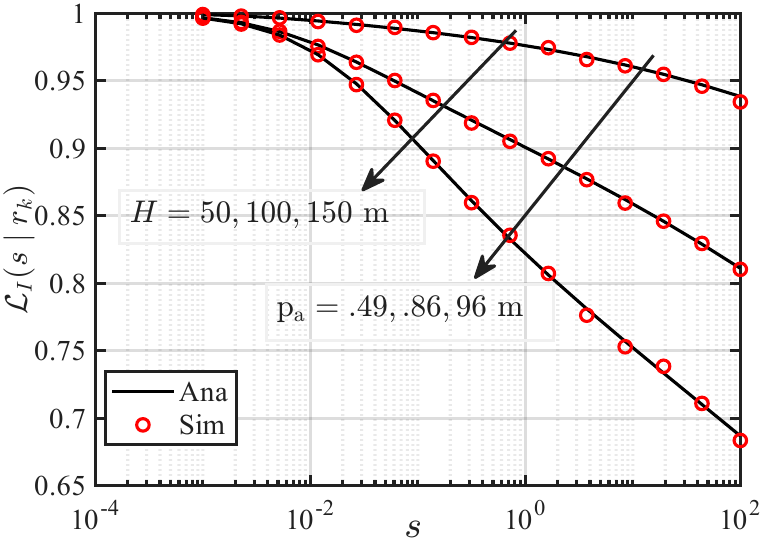}%
		\label{fig:LT_interference}}
	\hfil
	\subfloat[]{\includegraphics[width=0.45\columnwidth]{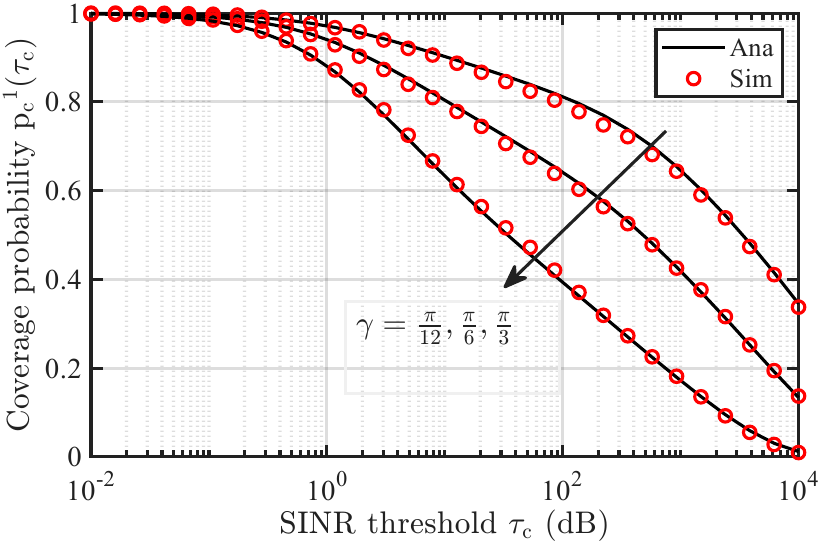}%
		\label{fig:coverage_3gamma}}
	\caption{(a)~LT of aggregate interference $\mathcal{L}_I(s \mid r_k)$ versus $s$ for UAV altitudes $H \in \{50, 100, 150\}$~m, with guard distance $r_k = 100$~\si{m}, beamwidth $\gamma = \pi/6$, and dense urban blockage parameters ($a = 9.61$, $b = 0.16$). (b)~CP ${\rm p_c}^1(\tau_{\mathrm{c}})$ versus SINR threshold $\tau_{\mathrm{c}}$\,(\si{dB}) for the nearest vehicle at $H=250$\,\si{m}, with beamwidth $\gamma\in\{\pi/12,\,\pi/6,\,\pi/3\}$. The sensing radius $r_\rms$ follows from the reliability constraint $\delta=0.5$, and interfering UAVs are thinned by ${\rm p_a}$.}
	\label{fig:LT_coverage_joint}
\end{figure}
\begin{figure}[t!]
	\centering
	\includegraphics[width=\linewidth]{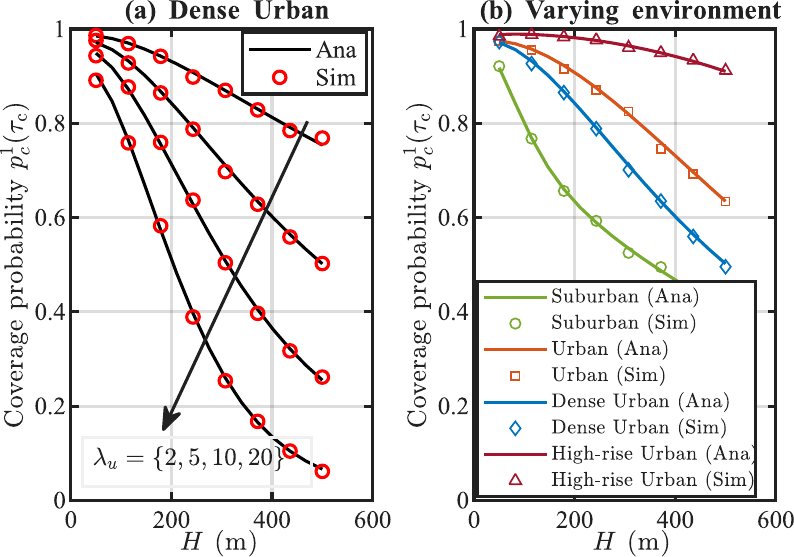}
	\caption{CP and UAV activation probability with UAV altitude $H$ for beamwidth. The sensing reliability threshold $\delta=0.2$ determines the critical sensing distance $d_s$ and corresponding horizontal sensing radius $r_{\rms}=\sqrt{d_\rms^2-H^2}$. The SINR threshold is fixed at $\tau_{\crm} = 10$ \si{dB}.}\label{fig:covprobwithh}
\end{figure}
\subsection{Analysis of the RC} Fig.~\ref{fig:ratecoveragev1}(a) illustrates the RC $r_c(\tau)$ of the nearest 
sensed vehicle versus UAV altitude $H$. At low altitudes, strong LoS interference 
dominates, while at large altitudes the signal strength deteriorates due to increased 
path loss. Higher UAV density consistently degrades RC through increased aggregate 
interference. Fig.~\ref{fig:ratecoveragev1}(b) shows that unlike RC, the typical RC 
peaks at a moderate altitude ($H \approx 75$--$100$~m) and decays monotonically 
beyond it. At low altitudes, few vehicles fall within the sensing region (reliability 
threshold $\delta = 0.2$), limiting ${\rm p_a}$. At large altitudes, both ${\rm p_a}$ 
and $r_c(\tau)$ decline simultaneously, compounding the degradation. In both panels, 
analytical results closely match Monte Carlo simulations, validating the derived 
framework. An optimal UAV altitude exists that maximizes typical rate coverage, and 
increasing UAV density reduces this peak.

\begin{figure}[t!]
	\centering
	\includegraphics[width=\linewidth]{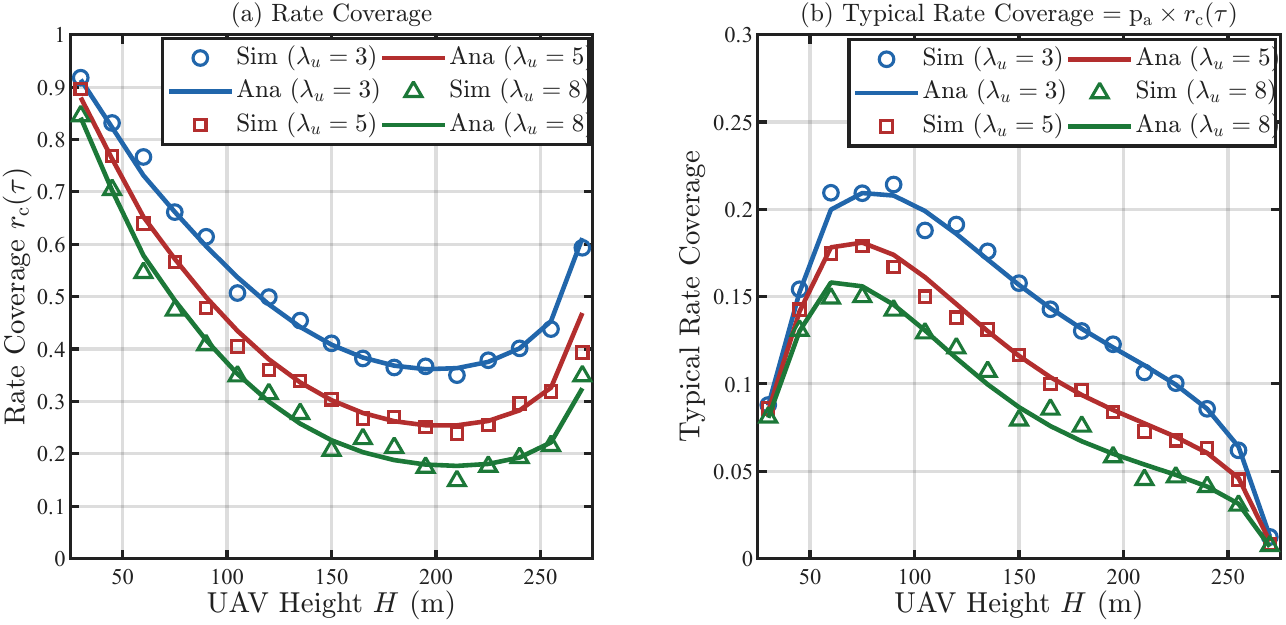}
	\caption{{(a) RC (b) the typical RC with UAV altitude for different UAV densities, with rate threshold $\tau=5$ \si{bps/Hz}.}}\label{fig:ratecoveragev1}
\end{figure}

\section{Conclusion}
This paper developed a tractable SG framework for sensing-assisted UAV networks serving urban vehicular communications. Roads were modeled as an MPLP, vehicles as MPLP-PPPs, and UAVs as a homogeneous 2D PPP at a fixed altitude operating under a sense-then-communicate protocol. Closed-form expressions were derived for the DP, sensing radius, sensing load distribution, UAV activation probability, CP, and RC, all validated by Monte Carlo simulations.

\noindent We observe that, at low altitude, the blockage-limited radius $r_{\rm L}$ dominates, creating a blockage-dominant regime in which increasing transmit power or antenna gain yields no improvement in sensing coverage. At higher altitudes, the SNR-limited radius $r_{\rm s}^{\rm snr}$ becomes the binding constraint as path loss and molecular absorption grow with the slant distance.  The crossover between these two regimes is governed jointly by the UAV altitude and the urban environment parameters, establishing an environment-dependent optimal altitude that maximizes the sensing radius. The UAV activation probability exhibits a non-monotonic dependence on altitude when blockage is present, first rising as the elevation angle improves LoS visibility and then falling as path loss dominates. In high-traffic conditions, the activation probability saturates, indicating that connectivity is ultimately limited by sensing geometry, particularly road density and sensing radius, rather than vehicular density. From the communication perspective, narrower beams improve both coverage and RC by simultaneously increasing directional gain and suppressing interference through the $\gamma/(2\pi)$ thinning factor. RC exhibits a non-monotonic behavior with altitude, while the typical RC  reveals a clearly optimal deployment altitude. High-rise urban environments maintain better coverage at large altitudes due to blockage-assisted interference suppression as compare to other environment.

\noindent These results collectively establish that UAV altitude and antenna beamwidth are tightly coupled design variables that cannot be optimized independently. Efficient sensing-assisted UAV networks for urban vehicular communications require their joint design. As future extensions of this work, joint optimization of UAV altitude, beamwidth, and deployment density to maximize a unified sensing-communication utility under practical constraints is a natural next step. Incorporating UAV and vehicle mobility to study the temporal evolution of the sensing-communication trade-off and extending the framework to heterogeneous multi-tier aerial networks are further directions of interest. 

%
%



%

\appendices
\section{}\label{proofthm1}
Let a vehicle be located  $\x\equiv(r,\theta)$ with respect to the typical
UAVs located at the origin. 		
A vehicle is detected if the sensing SNR exceeds the threshold $\tau_\rms$. Since detection requires both angular alignment and a LoS path for the radar round-trip,
we condition on $\{|\theta-{\Theta_{\ob}}|\le \gamma/2\}$ and the LoS event, yielding
\begin{align*}
	{\rm P_d}(r)
	&=
	\mathbb{P}\!\left(
	|\theta-{\Theta_{\ob}}|\le \gamma/2
	\right)\;p_{\sf L}(r,H)\\
	&\quad\times\mathbb{P}\!\left(
	\mathrm{SNR}_\rms(d)\ge \tau_\rms
	\,\big|\,
	|\theta-{\Theta_{\ob}}|\le \gamma/2,\,\text{LoS}
	\right).
\end{align*}\noindent Because $\theta$ is uniformly distributed over $[0,2\pi)$, the angular alignment
probability is $	\mathbb{P}\!\left(
|\theta-{\Theta_{\ob}}|\le \gamma/2
\right)
=
\frac{\gamma}{2\pi}.$ The LoS probability $p_{\sf L}(r,H)$ is given by \eqref{eq:PLoS}. Conditioned on angular alignment and LoS, the DP is
\begin{align*}
	\mathbb{P}\!\left(
	\mathrm{SNR}_\rms(r)\ge \tau_{\rms}
	\right)
	&=
	\mathbb{P}\!\left(
	\sigma \ge
	\frac{
		\tau_{\rms} (4\pi)^3 d^{2\alpha} N_0
		\exp\!\left(2\kappa_f d\right)
	}{
		P_\trm G_s^2(\gamma) \lambda_c^2
	}
	\right).
\end{align*}Simplifying further completes the proof.
\section{}\label{proofthm2}
The sensing radius satisfies $r_{\rm s}=\min(r_{\rm s}^{\rm snr},\,r_{\rm L})$, where $r_{\rm L}=H/\tan(\pi a/180)$ is the blockage-limited radius from the step-function approximation~\eqref{eq:PLoS_step}. Under this approximation, $p_{\sf L}(r,H)=1$ for $r\le r_{\rm L}$, so the conditional DP reduces to the SNR term alone
\begin{align*}
	{{\rm P_d^{cond}}(r)}
	=\exp\!\left(
	-\frac{
		\tau_{\rms} (4\pi)^3 (d(r))^{2\alpha} N_0
		\exp\!\left(2\kappa_f d(r)\right)
	}{
		\trp G_s^2(\gamma) \lambda_c^2
	}
	\right).
\end{align*}The constraint ${\rm P_d^{cond}}(r)\ge\delta$ requires both the exponential SNR term and $p_{\sf L}(r,H)$ to be sufficiently large. Considering the SNR constraint alone and imposing ${\rm P_d^{cond}}(r)\ge \delta$ yields
\begin{align*}
	\frac{
		\tau_\rms (4\pi)^3 (d(r))^{2\alpha} N_0
		\exp\!\left(2\kappa_f d(r)\right)
	}{
		\trp G_s^2(\gamma) \lambda_c^2
	}
	&\le
	-\ln(\delta)\\
	(d(r))^{2\alpha}
	\exp\!\left(2\kappa_f d(r)\right)
	&\le
	\frac{
		-\ln(\delta)\, \trp G_\rms^2(\gamma) \lambda_c^2
	}{
		\tau_\rms (4\pi)^3 N_0
	}.
\end{align*}Simplifying further and applying the definition of the
Lambert--${\rm W}$ function, the power-limited maximum sensing distance $d_{\rm s}$ is obtained
as
\begin{align*}
	d_{\rm s}=d(r^{\rm SNR}_{\rm s})
	=
	\frac{\alpha}{\kappa_f}
	\,{\rm W}\!\left(
	\frac{\kappa_f}{\alpha}
	\left(
	\frac{
		-\ln(\delta)\, \trp G_s^2(\gamma) \lambda_c^2
	}{
		\tau_\rms (4\pi)^3 N_0
	}
	\right)^{\frac{1}{2\alpha}}
	\right).
\end{align*}Hence, the power-limited horizontal radius is $r_{\rm s}^{\rm snr}
=
\sqrt{
	d_{\rm s}^2 - H^2
}$. Combining both constraints yields $r_{\rm s}=\min(r_{\rm s}^{\rm snr},\, r_{\rm L})$.
\section{}\label{proofthmpmf}
The typical UAV is located at the origin with its directional antenna oriented along the $+\sfx$ direction ($\Theta_{\ob}=0$).	The number of detectable vehicles $N(r)$ falling in the sensing zone of the UAV is $N(r_{\rm s}) = N_{\rm h}(r_{\rm s}) + N_{\rm v}(r_{\rm s}),$ where $N_{\rm h}(r_{\rms})$ and $N_{\rmv}{(r_{\rms})}$ denotes the numbers of detectable vehicles located on roads parallel and perpendicular to $\sfx$ axis, respectively. 
Due to independence, the PGF of $N(r_{\rms})$ is the product of the PGFs of $N_{\rm h}(r_{\rms})$ and $N_{\rm v}(r_{\rms})$. As illustrated in Fig. \ref{fig:proofdiagram}(A), the length of the projection of the arc on the $\sfy$ axis is $2 r_{\rms}\sin(\gamma/2)$. Hence, the number of roads parallel to $\sfx$ axis intersecting the sensing region is Poisson distributed with mean $2\lambda_{\ell} r_{\rms}\sin(\gamma/2)$. On each road at distance $\rho$, vehicles form a 1D PPP with density $\lambda$. Since only LoS vehicles are detectable, we thin each vehicle independently with retention probability $p_{\sf L}(\sqrt{x^2+\rho^2},H)$. The retained process is an inhomogeneous PPP with mean count $\Lambda_{\rm h}(\rho)=\int_{\ell_\rmh(\rho,\gamma)}\lambda\,p_{\sf L}(\sqrt{x^2+\rho^2},H)\,\dv x$. The PGF for the number of detectable vehicles conditioned on a road at distance $\rho$ is
$f_{\rm h}(\rho,s)
=
\exp\!\big(
\Lambda_\rmh(\rho)\,(s-1)
\big)$. Conditioned on the event that exactly $n$ roads parallel to $\sfx$ axis intersect the sensing
sector, and using the independence of different roads, the conditional PGF is
\[
\PGF_{N_{\rmh}(r_{\rms})\mid n}(s)
=
\left(
\frac{1}{r_{\rms}\sin(\gamma/2)}
\int_{0}^{r_{\rms}\sin(\gamma/2)}
f_{\rm h}(\rho,s)\,
\mathrm{d}\rho
\right)^{n}.
\]Deconditioning with respect to the Poisson-distributed number of intersecting
horizontal roads yields
\begin{align*}
	&\PGF_{N_{\rm h}}(s)
	\stackrel{(a)}=
	\sum\nolimits_{n=0}^{\infty}
	\frac{e^{-2\lambda_\ell r_{\rms}\sin(\gamma/2)}
		\left(2\lambda_\ell r_{\rms}\sin(\gamma/2)\right)^n}{n!}
	\PGF_{N_{\rm h}\mid n}(s)
	\\
	&\stackrel{(b)}=
	\exp\left(
	2\lambda_\ell\int_{0}^{r_{\rms}\sin(\gamma/2)}\left(e^{\Lambda_\rmh(\rho)(s-1)}
	-1\right)\dv\rho\right),
\end{align*}where $(a)$ is after deconditioning using the Poisson distribution and step $(b)$ obtained by simplifying the step $(a)$.
\begin{figure}[t!]
	\centering
	\includegraphics[width=.78\linewidth]{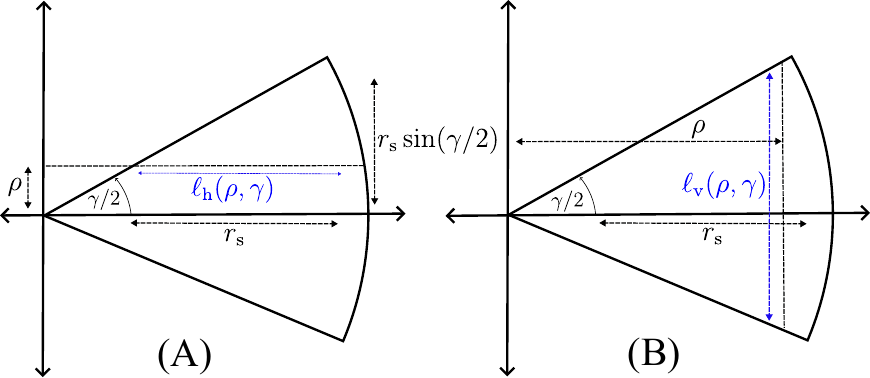}
	\caption{Diagram illustrating the (A) horizontal and the (B) vertical chord located at a distance $\rho$ from origin.}
	\label{fig:proofdiagram}
\end{figure}Similarly, using the Fig. \ref{fig:proofdiagram}(B) we can determine the length of the $\ell_\rmv(\rho,\gamma)$. Then using the similar steps we can find the PGF for $N_{\rmv}(r_{\rms})$.
\section{}\label{proofLT}
Using the definition LT $\mathcal{L}_{I_k}(s)$ of $I_{k}$ is
\begin{align}
	&\stackrel{(a)}{=}
	\mathbb{E}_{\Phi_u^{\rm a}}
	\!\left[
	\prod\nolimits_{\y \in \Phi_u^{\rm a}\setminus \bt_{2}(\x_k,r_k)}\right.\notag\\
	&\quad\quad\left.
	\mathbb{E}_{h}
	\!\left(
\expS{- s P_{\trm} G_{\y} h
	(d(x))^{-\alpha}
	e^{-\kappa_f d(x)}\1(\text{LoS})}
	\right)
	\right] \notag\\
	&\stackrel{(b)}{=}
	\mathbb{E}_{\Phi_u^{\rm a}}
	\!\left[
	\prod\nolimits_{\y \in \Phi_u^{\rm a}\setminus \bt_{2}(\x_k,r_k)}\right.\nonumber\\
	&\quad\left.
	\left(1-p_{\sf L}(x,H)\frac{
		s P_{\trm} G_{\y}
		(d(x))^{-\alpha}
		e^{-\kappa_f d(x)}
	}{
		1+
		s P_{\trm} G_{\y}
		(d(x))^{-\alpha}
		e^{-\kappa_f d(x)}
	}\right)
	\right] \notag\\
&\stackrel{(c)}{=}
\exp\!\left(
- ({\gamma\lambda_u {\rm p_a}}/{(2\pi)})
\int_{\mathbb{R}^2\setminus \bt_{2}(\ob,r_k)}
p_{\sf L}(x,H)\right.\notag\\
&\quad\quad\quad\quad\left.\times\frac{
	s P_{\trm} G_{\y}
	(d(y))^{-\alpha}
	e^{-\kappa_f d(y)}
}{
	1+
	s P_{\trm} G_{\y}
	(d(y))^{-\alpha}
	e^{-\kappa_f d(y)}
}\dv \y
\right) \notag\\
&\stackrel{(d)}{=}
\exp\!\left(
-  \gamma \lambda_u {\rm p_{a}}
\int_{r_k}^{\infty}
\frac{
	s P_\trm G_{\y} (d(x))^{-\alpha}
	e^{-\kappa_f d(x)}p_{\sf L}(x,H)
}{
	1+
	s P_\trm G_{\y} (d(x))^{-\alpha}
	e^{-\kappa_f d(x)}
}\,x \dv x\right)\notag
\end{align}where step $(a)$ follows from the independence of Rayleigh fading coefficients
$\{h_{\y}\}$ across different UAV--vehicle links, conditioned on $\Phi_u^{\rm a}$.
Step $(b)$ is obtained by averaging the exponential fading MGF over the LoS/NLoS
states. Step $(c)$ follows from the substitution $\y \mapsto \y-\x_k$ and an
application of the PGFL of a homogeneous PPP. Step $(d)$ is obtained by converting
the resulting two-dimensional integral into polar coordinates. Simplifying further
completes the proof.
\section{}\label{proof}
 The conditional CP $\mathbb{P}(\SINR_k>\tau \mid R_k=r)$ is
\begin{align}
	&=
	\mathbb{P}\!\left(
	h_{\ob}
	>
	\frac{\tau (I_k+N_{0}) \left(d(r)\right)^{\alpha} e^{\kappa_f d(r)}}
	{P_\trm G_\crm}
	\right) \notag\\
	&\stackrel{(a)}{=}
	\mathbb{E}_{I_k}
	\!\left[
	\exp\!\left(
	-\frac{\tau (I_k+N_{0}) \left(d(r)\right)^{\alpha} e^{\kappa_f d(r)}}
	{P_\trm G_\crm}
	\right)
	\right] \notag\\
	&\stackrel{(b)}{=}
e^{	-\frac{\tau N_{0} \left(d(r)\right)^{\alpha} e^{\kappa_f d(r)}}
	{P_\trm G_\crm}}
	\mathbb{E}_{I_k}
	\!\left[
	\exp\!\left(
	-\frac{\tau \left(d(r)\right)^{\alpha} e^{\kappa_f d(r)}I_k}{P_\trm G_\crm}
	\right)
	\right] \notag\\
	&\stackrel{(c)}{=}
e^{	-\frac{\tau N_{0} \left(d(r)\right)^{\alpha} e^{\kappa_f d(r)}}
	{P_\trm G_\crm}}
	\mathcal{L}_{I_k}\!\left(
	\frac{\tau \left(d(r)\right)^{\alpha} e^{\kappa_f d(r)}}{P_\trm G_\crm}
	\right),
\end{align}where step $(a)$ is obtained using the CDF of the RV $h_{\ob}$, $(b)$ is obtained by separating the noise and the interference and finally $(c)$ is obtained using the LT of the interference $I_{k}$. Deconditioning over $R_k$ and normalizing by
$\mathbb{P}(R_k \le r_{\rms})$ yields the desired result.
		
\section{}\label{rcproof}
{The per-sector load~$L$ determines the bandwidth share~$B/L$ allocated to each vehicle. The} RC must be conditioned on the event
$N(r_{\mathrm{s}})\ge 1$. Using the law of total probability over the
{per-sector} sensing load, we write
\begin{align}
	&{r}_\crm(\tau)=
	\mathbb{P}\!\left(
	({B}/{L})\log_2\!\left(1+\SINR\right) > \tau
	\;\middle|\;
	N(r_{\mathrm{s}})\ge 1
	\right) \nonumber\\
	&\stackrel{(a)}=
	\sum\nolimits_{k=1}^{\infty}
	\mathbb{P}\!\left(\mathcal{R}_k>\tau \mid N(r_{\mathrm{s}})=k\right)\mathbb{P}\!\left(N(r_{\mathrm{s}})=k \mid N(r_{\mathrm{s}})\ge 1\right)\notag\\
	&\stackrel{(b)}=\sum\nolimits_{k=1}^{\infty}\mathbb{P}\!\left(
	\mathrm{SINR}_k > 2^{\frac{\tau k}{B}}-1
	\right)\times	({p_k(r_{\mathrm{s}})}/{(1-p_0(r_{\mathrm{s}}))})\notag,
\end{align}where $(a)$ is obtained by deconditioning the conditional RC and $(b)$ is obtained by simplifying the step $(a)$ and replacing the conditional load distribution as $	\mathbb{P}\!\left(N(r_{\mathrm{s}})=k \mid N(r_{\mathrm{s}})\ge 1\right)
=
{p_k(r_{\mathrm{s}})}/{(1-p_0(r_{\mathrm{s}}))}.$ Finally, using \eqref{CP} completes the proof. {Using similar steps and conditioning we can derive the RC $r_{\crm}^{k}(\tau)$ for the $k$th-nearest vehicle.}


\ifCLASSOPTIONcaptionsoff
  \newpage
\fi



%
\bibliographystyle{IEEEtran}
\bibliography{scibib,NewCompPaperDB}

\begin{thebibliography}{10}
\providecommand{\url}[1]{#1}
\csname url@samestyle\endcsname
\providecommand{\newblock}{\relax}
\providecommand{\bibinfo}[2]{#2}
\providecommand{\BIBentrySTDinterwordspacing}{\spaceskip=0pt\relax}
\providecommand{\BIBentryALTinterwordstretchfactor}{4}
\providecommand{\BIBentryALTinterwordspacing}{\spaceskip=\fontdimen2\font plus
\BIBentryALTinterwordstretchfactor\fontdimen3\font minus
  \fontdimen4\font\relax}
\providecommand{\BIBforeignlanguage}[2]{{%
\expandafter\ifx\csname l@#1\endcsname\relax
\typeout{** WARNING: IEEEtran.bst: No hyphenation pattern has been}%
\typeout{** loaded for the language `#1'. Using the pattern for}%
\typeout{** the default language instead.}%
\else
\language=\csname l@#1\endcsname
\fi
#2}}
\providecommand{\BIBdecl}{\relax}
\BIBdecl

\bibitem{shi2018drone}
W.~Shi, H.~Zhou, J.~Li, W.~Xu, N.~Zhang, and X.~Shen, ``Drone assisted
  vehicular networks: Architecture, challenges and opportunities,'' \emph{IEEE
  Netw.}, vol.~32, no.~3, pp. 130--137, 2018.

\bibitem{UAVassisted}
M.~Khabbaz, J.~Antoun, and C.~Assi, ``Modeling and performance analysis of
  {U}{A}{V}-assisted vehicular networks,'' \emph{IEEE Trans. Veh. Technol.},
  vol.~68, no.~9, pp. 8384--8396, 2019.

\bibitem{huang2024_low}
C.~Huang, S.~Fang, H.~Wu, Y.~Wang, and Y.~Yang, ``Low-altitude intelligent
  transportation: System architecture, infrastructure, and key technologies,''
  \emph{J. Ind. Inf. Integr.}, vol.~42, p. 100694, Nov. 2024.

\bibitem{jiang2024_6g_ntn_lae}
Y.~Jiang \emph{et~al.}, ``6{G} non-terrestrial networks enabled low-altitude
  economy: Opportunities and challenges,'' \emph{arXiv preprint
  arXiv:2311.09047}, 2024.

\bibitem{alhourani2014modeling}
A.~Al-Hourani, S.~Kandeepan, and A.~Jamalipour, ``Modeling air-to-ground path
  loss for low altitude platforms in urban environments,'' in \emph{Proc. IEEE
  Glob. Commun. Conf. (GLOBECOM)}, 2014, pp. 2898--2904.

\bibitem{wu2018}
Q.~Wu, Y.~Zeng, and R.~Zhang, ``Joint trajectory and communication design for
  multi-uav enabled wireless networks,'' \emph{IEEE Trans. Wireless Commun.},
  vol.~17, no.~3, pp. 2109--2121, Mar. 2018.

\bibitem{deng2019joint}
L.~Deng, G.~Wu, J.~Fu, Y.~Zhang, and Y.~Yang, ``Joint resource allocation and
  trajectory control for {U}{A}{V}-enabled vehicular communications,''
  \emph{IEEE Access}, vol.~7, pp. 132\,806--132\,815, 2019.

\bibitem{He2019AntiInterference}
Y.~He, D.~Zhai, R.~Zhang, X.~Du, D.~Zhou, and M.~Guizani, ``An
  anti-interference scheme for uav information link in air-ground integrated
  vehicular networks,'' \emph{Sensors}, vol.~19, no.~21, p. 4742, 2019.

\bibitem{Oubbati2019ECaD}
O.~S. Oubbati, M.~Mozaffari, N.~Chaib, P.~Lorenz, M.~Atiquzzaman, and
  A.~Jamalipour, ``E{C}a{D}: Energy-efficient routing in flying ad hoc
  networks,'' \emph{Int. J. Commun. Syst.}, vol.~32, no.~18, p. e4156, 2019.

\bibitem{Hu2026MADQN}
B.~Hu, H.~Liu, J.~Du, M.~L{\'o}pez-Ben{\'i}tez, C.~Wu, X.~Chu, and D.~Niyato,
  ``{M}{A}{D}{Q}{N}-enhanced computation offloading and resource allocation for
  6{G} low-altitude economy vehicular networks,'' \emph{IEEE Trans. Cogn.
  Commun. Netw.}, vol.~12, pp. 2603--2617, 2026.

\bibitem{Pang2024DynamicISAC}
X.~Pang, S.~Guo, J.~Tang, N.~Zhao, and N.~Al-Dhahir, ``Dynamic isac beamforming
  design for {U}{A}{V}-enabled vehicular networks,'' \emph{IEEE Trans. Wireless
  Commun.}, vol.~23, no.~11, pp. 16\,852--16\,864, 2024.

\bibitem{Liu2022ISAC}
F.~Liu, Y.~Cui, C.~Masouros, J.~Xu, W.~Zhang, X.~Gao, and J.~An, ``Integrated
  sensing and communications: Toward dual-functional wireless networks for 6{G}
  and beyond,'' \emph{IEEE J. Sel. Areas Commun.}, vol.~40, no.~6, pp.
  1728--1767, 2022.

\bibitem{Shen2021ISAC}
H.~Shen, W.~Xu, S.~Jin, C.~Wen, and X.~You, ``Integrated sensing and
  communication: A survey,'' \emph{IEEE Commun. Surveys Tuts.}, vol.~23, no.~4,
  pp. 2404--2437, 2021.

\bibitem{armeniakos2025stochastic}
H.~K. Armeniakos, P.~S. Bithas, S.~A. Tegos, A.~G. Kanatas, and G.~K.
  Karagiannidis, ``Stochastic geometry for modeling and analysis of sensing and
  communications: A survey,'' \emph{IEEE Commun. Surveys Tuts.}, vol.~28, pp.
  2691--2724, 2026.

\bibitem{zeng2019accessing}
Y.~Zeng, Q.~Wu, and R.~Zhang, ``Accessing from the sky: A tutorial on {U}{A}{V}
  communications for 5{G} and beyond,'' \emph{Proc. IEEE}, vol. 107, no.~12,
  pp. 2327--2375, 2019.

\bibitem{Zhang2019MmWaveUAV}
L.~Zhang, H.~Zhao, S.~Hou, Z.~Zhao, H.~Xu, X.~Wu, Q.~Wu, and R.~Zhang, ``A
  survey on 5{G} millimeter wave communications for {U}{A}{V}-assisted wireless
  networks,'' \emph{IEEE Access}, vol.~7, pp. 117\,460--117\,504, 2019.

\bibitem{Mozaffari2019}
M.~Mozaffari, W.~Saad, M.~Bennis, and M.~Debbah, ``A tutorial on {U}{A}{V}s for
  wireless networks: Applications, challenges, and open problems,'' \emph{IEEE
  Communications Surveys \& Tutorials}, vol.~21, no.~3, pp. 2334--2360, 2019.

\bibitem{9508366}
A.~Baltaci, E.~Dinc, M.~Ozger, A.~Alabbasi, C.~Cavdar, and D.~Schupke, ``A
  survey of wireless networks for future aerial communications
  ({F}{A}{C}{O}{M}),'' \emph{IEEE Commun. Surv. Tutor.}, vol.~23, no.~4, pp.
  2833--2884, 2021.

\bibitem{telikani2025unmanned}
A.~Telikani, A.~Sarkar, B.~Du, F.~Santoso, J.~Shen, J.~Yan, J.~Yong, and E.~W.
  Yap, ``Autonomous aerial vehicles-aided intelligent transportation systems:
  Vision, challenges, and opportunities,'' \emph{IEEE Commun. Surv. Tutor.},
  vol.~27, no.~6, pp. 3772--3819, 2025.

\bibitem{he2023aerial}
Y.~He, D.~Wang, F.~Huang, R.~Zhang, and L.~Min, ``Aerial-ground integrated
  vehicular networks: A {U}{A}{V}-vehicle collaboration perspective,''
  \emph{IEEE Trans. Intell. Transp. Syst.}, vol.~25, no.~6, pp. 5154--5169,
  2023.

\bibitem{zhang2024performance}
B.~Zhang, Z.~He, Y.~Feng, and Z.~Han, ``Performance analysis and 3d position
  deployment for {V}2{V}-assisted {U}{A}{V} communications in vehicular
  networks,'' \emph{IEEE Trans. Veh. Tech.}, vol.~73, no.~12, pp.
  19\,361--19\,373, 2024.

\bibitem{jiang2024performance}
L.~Jiang, Q.~Sun, H.~Chen, Y.~Sun, Y.~Cao, H.~Yu, H.~Li, X.~Zhao, Y.~Zhao, and
  S.~Wang, ``Performance analysis of the air-ground integrated vehicular
  networks: A hierarchical model approach,'' \emph{Ad Hoc Netw.}, vol. 154, p.
  103359, 2024.

\bibitem{chetlur2018coverage}
V.~V. Chetlur and H.~S. Dhillon, ``Coverage analysis of a vehicular network
  modeled as {C}ox process driven by {P}oisson line process,'' \emph{IEEE
  Trans. Wireless Commun.}, vol.~17, no.~7, pp. 4401--4416, 2018.

\bibitem{31}
X.~Jing, F.~Liu, C.~Masouros, and Y.~Zeng, ``{I}{S}{A}{C} from the sky:
  {U}{A}{V} trajectory design for joint communication and target
  localization,'' \emph{IEEE Trans. Wireless Commun.}, vol.~23, no.~10, pp.
  12\,857--12\,872, 2024.

\bibitem{32}
Z.~Lyu, G.~Zhu, and J.~Xu, ``Joint maneuver and beamforming design for
  {UAV}-enabled integrated sensing and communication,'' \emph{IEEE Trans.
  Wireless Commun.}, vol.~22, no.~4, pp. 2424--2440, Apr. 2023.

\bibitem{33}
K.~Meng, Q.~Wu, S.~Ma, W.~Chen, K.~Wang, and J.~Li, ``Throughput maximization
  for {UAV}-enabled integrated periodic sensing and communication,'' \emph{IEEE
  Trans. Wireless Commun.}, vol.~22, no.~1, pp. 671--687, Jan. 2023.

\bibitem{chetlur2020shortest}
V.~V. Chetlur, H.~S. Dhillon, and C.~P. Dettmann, ``Shortest path distance in
  {M}anhattan {P}oisson line cox process,'' \emph{J. Stat. Phys.}, vol. 181,
  no.~6, pp. 2109--2130, 2020.

\bibitem{dhillon2022poisson}
H.~S. Dhillon and V.~V. Chetlur, ``The {P}oisson {L}ine {C}ox process,'' in
  \emph{{P}oisson Line Cox Process: Foundations and Applications to Vehicular
  Networks}.\hskip 1em plus 0.5em minus 0.4em\relax Springer, 2022, pp. 31--46.

\bibitem{haenggibook}
M.~Haenggi, \emph{Stochastic geometry for wireless networks}.\hskip 1em plus
  0.5em minus 0.4em\relax Cambridge University Press, 2012.

\bibitem{swerling2003probability}
P.~Swerling, ``Probability of detection for fluctuating targets,'' \emph{IRE
  Trans. Inf. Theory}, vol.~6, no.~2, pp. 269--308, 2003.

\bibitem{Persson2017RadarRCS}
B.~Persson, ``Radar target modeling using in-flight radar cross-section
  measurements,'' \emph{J. Aircr.}, vol.~54, no.~1, pp. 284--291, Jan. 2017.

\bibitem{Fang2020StochasticRadarRCS}
Z.~Fang, Z.~Wei, X.~Chen, H.~Wu, and Z.~Feng, ``Stochastic geometry for
  automotive radar interference with {RCS} characteristics,'' \emph{IEEE
  Wireless Commun. Lett.}, vol.~9, no.~11, pp. 1817--1820, Nov. 2020.

\bibitem{ITU-R-P676}
P.~Series, ``Attenuation by atmospheric gases and related effects,'' Tech.
  Rep., 2019.

\bibitem{al2014optimal}
A.~Al-Hourani, S.~Kandeepan, and S.~Lardner, ``Optimal {L}{A}{P} altitude for
  maximum coverage,'' \emph{IEEE Wireless Commun. Lett.}, vol.~3, no.~6, pp.
  569--572, 2014.

\bibitem{balanis2016}
C.~A. Balanis, \emph{Antenna Theory: Analysis and Design}, 4th~ed.\hskip 1em
  plus 0.5em minus 0.4em\relax Wiley, 2016.

\bibitem{skolnik2008}
M.~I. Skolnik, \emph{Radar Handbook}, 3rd~ed.\hskip 1em plus 0.5em minus
  0.4em\relax McGraw-Hill, 2008.

\bibitem{nabil2024beamwidth}
Y.~Nabil, H.~ElSawy, S.~Al-Dharrab, H.~Mostafa, and H.~Attia, ``Beamwidth
  design tradeoffs in radar-aided millimeter-wave cellular networks: A
  stochastic geometry approach,'' \emph{IEEE Access}, vol.~12, pp.
  26\,196--26\,211, 2024.

\bibitem{ram2025radar}
A.~Sneh and S.~S. Ram, ``Radar operating metrics and network throughput for
  integrated sensing and communications in millimeter-wave urban
  environments,'' in \emph{Proc. RadarConf24}.\hskip 1em plus 0.5em minus
  0.4em\relax IEEE, 2024, pp. 1--6.

\bibitem{liu2018mu}
F.~Liu, C.~Masouros, A.~Li, H.~Sun, and L.~Hanzo, ``{M}{U}-{M}{I}{M}{O}
  communications with {M}{I}{M}{O} radar: From co-existence to joint
  transmission,'' \emph{IEEE Trans. Wireless Commun.}, vol.~17, no.~4, pp.
  2755--2770, 2018.

\bibitem{baccelli1997stochastic}
F.~Baccelli, M.~Klein, M.~Lebourges, and S.~Zuyev, ``Stochastic geometry and
  architecture of communication networks,'' \emph{Telecommunication Systems},
  vol.~7, no.~1, pp. 209--227, 1997.

\end{thebibliography}
\vspace{12pt}

%





\end{document}